\documentclass[final,conference,twocolumn,letterpaper]{IEEEtran}
\usepackage{fancyhdr}
\usepackage{graphicx}
\usepackage[utf8]{inputenc}

\graphicspath{{figures/}}

\usepackage{url}
\usepackage{wrapfig}
\usepackage{units}

\usepackage[printonlyused]{acronym}

\usepackage{calc}
\usepackage{graphicx}
\usepackage[lined,boxed, vlined,ruled,linesnumbered]{algorithm2e}
\usepackage{amsmath,amssymb,amsthm,amsfonts,amsbsy}

\usepackage{algpseudocode}
\usepackage{color}
\usepackage[table]{xcolor} 
\usepackage[caption=false]{subfig}
\usepackage{dblfloatfix}
\usepackage{nicefrac}
\usepackage{tcolorbox}
\usepackage{txfonts}

\usepackage{multirow}

\newcolumntype{C}[1]{>{\centering\arraybackslash}p{#1}}

\usepackage{booktabs}

\definecolor{BgGray}{gray}{0.7}%
\definecolor{BgGray2}{gray}{0.96}%
\definecolor{RowColorOdd}{named}{BgGray2}%
\definecolor{RowColorEven}{named}{white}%
\definecolor{comments}{gray}{.5}
\definecolor{Gray}{gray}{0.85}

\newcommand\colheadbegin{\hline\rowcolor{BgGray}}
\newcommand\colheadend{\\\hline}

\usepackage{pifont}
\usepackage[multiuser,nomargin,marginclue,footnote]{fixme}
\fxusetheme{color}
\fxsetup{targetlayout=colorcb}
\FXRegisterAuthor{all}{anall}{ALL}
\FXRegisterAuthor{md}{anmd}{MD}
\FXRegisterAuthor{tolja}{antolja}{TOLJA}
\FXRegisterAuthor{pg}{anpg}{PG}
\FXRegisterAuthor{mch}{anmc}{MC}
\FXRegisterAuthor{cp}{ancp}{CP}
\FXRegisterAuthor{awo}{anawo}{AWO}

\fxsetup{status=draft}

\begin{document}
\providetoggle{techreport}
\settoggle{techreport}{false}

\title{LtFi: Cross-technology Communication for RRM between LTE-U and IEEE 802.11}

\author{
\IEEEauthorblockN{Piotr Gawłowicz, Anatolij Zubow and Adam Wolisz}
\IEEEauthorblockA{\{gawlowicz, zubow, wolisz\}@tkn.tu-berlin.de}
Technische Universität Berlin, Germany\\
}

\maketitle


\begin{abstract}
\iftoggle{techreport}{
\begin{center}
\begin{minipage}[t]{0.7\textwidth}
}%
{
}
Cross-technology communication (CTC) was proposed in recent literature as a way to exploit the opportunities of collaboration between heterogeneous wireless technologies. This paper presents LtFi, a system which enables to set-up a CTC between nodes of co-located LTE-U and WiFi networks. LtFi follows a two-step approach: using the air-interface LTE-U BSs are broadcasting connection and identification data to adjacent WiFi nodes, which is used to create a bi-directional control channel over the wired Internet. This way LtFi enables the development of advanced cross-technology interference and radio resource management schemes between heterogeneous WiFi and LTE-U networks. 

LtFi is of low complexity and fully compliant with LTE-U technology and works on WiFi side with COTS hardware. It was prototypically implemented and evaluated. Experimental results reveal that LtFi is able to reliably decoded the data transmitted over the LtFi air-interface in a crowded wireless environment at even very low LTE-U receive power levels of -92\,dBm. Moreover, results from system-level simulations show that LtFi is able to accurately estimate the set of interfering LTE-U BSs in a typical LTE-U multi-cell environment.

\iftoggle{techreport}{
\end{minipage}
\end{center}
}%
{
}
\end{abstract}



\section{Introduction}

It is expected that the proliferation of WiFi and the appearance of new communication technologies like LTE in unlicensed (i.e. LTE-U~\cite{lteu_forum} and LTE-LAA~\cite{3gpp}) combined with exponential growth in the number of such devices will results in performance degradation in both WiFi and LTE-U/LAA networks operating in 5\,GHz ISM band due to interference in dense environments~\cite{cisco_forecast}. Although LTE-U/LAA and WiFi technologies have similar physical layers they are unable to decode each other’s packets and have to rely on energy-based carrier sensing for co-existence. However, an efficient coordination would require enabling direct cross-technology communication (CTC) between them.

Potentially LTE-U/LAA BSs and WiFi AP are connected over the backhaul to the Internet and hence are able to communicate with each other. Unfortunately, this possibility cannot be directly utilized as a discovery component for detection and identification of co-located and interfering LTE-U and WiFi cells is missing. CTC approaches known from literature are focused on scenarios of WiFi and sensor network technologies like ZigBee~\cite{cmorse,emf}. However, they are not directly applicable to our LTE-U/WiFi scenario.

We present LtFi, an approach which enables to establish a CTC between the nodes of LTE-U and WiFi networks. Therefore, LtFi exploits the option of subframe puncturing or Almost Blank Subframe (ABSF) insertion into LTE-U's air interface, whose relative position is used to modulate information data on the CTC channel. On WiFi side we exploit the capabilities of COTS HW --- monitoring the MAC state (signal detection logic) of the WiFi NIC allowing us to distinguish between WiFi and non-WiFi signals. The latter is used as input for the CTC demodulator. The so established unidirectional over-the-air CTC channel enables the LTE-U network to communicate connection and identification information, i.a. public IP address of the LtFi Management Unit (LtFiMU), to co-located WiFi APs. The WiFi APs in turn use that information to establish a secure, point-to-point control channel between each pair of WiFi AP and corresponding LtFiMU which acts as a proxy towards the LTE-U network over the wired backbone Internet. 

The so established bidirectional control channel can be used by various applications considering interference and radio resource management, e.g. load-aware adaptation of LTE-U CSAT cycle and cross-technology aware MAC backoff schemes, to optimize the operation of co-located LTE-U and WiFi networks.

\medskip

\noindent \textbf{Contributions:} 
LtFi is the first system that allows cross-technology communication between LTE-U and WiFi. LtFi is fully compliant and transparent with LTE-U technology. On WiFi side LtFi works with COTS hardware and requires only a simple software installation process at APs without modification within the WiFi STAs. LtFi provides a fine-grained cross-technology proximity detection mechanism using its air-interface enabling advanced interference and radio resource management schemes between WiFi and LTE-U.
LtFi was prototypically implemented and evaluated. Results reveal that the data rate achieved over the LtFi air-interface is between 50 to 650\,bps which is sufficient for transmission of control data.
The wireless CTC signal can be decoded at low receive signal power levels, i.e. -92\,dBm. The results from network simulations show that LtFi is able to accurately estimate the set of interfering LTE-U BSs in a typical LTE multi-cell environment.


\section{Background Knowledge} \label{lte_primer}

Since, neither LTE-U nor WiFi are able to decode each other’s frames, other coexistence mechanisms have to be applied. The current solutions rely on Listen before Talk (LBT) based on energy-based carrier sensing. This section gives a brief introduction into the relevant parts of the LTE-U and WiFi standards.

%
%
\subsection{LTE-U}

LTE-U is being specified by the LTE-U forum~\cite{lteu_forum} as first cellular solution for use of unlicensed band for the downlink (DL) traffic. The LTE carrier aggregation framework supports utilization of the unlicensed band as a secondary cell in addition to the licensed anchor serving as the primary cell~\cite{ni_lte_whitepaper}. The LTE-U channel bandwidth is set to 20\,MHz which corresponds to the smallest channel width in WiFi. LTE-U can be deployed in USA, China and India, where LBT is not required for unlicensed channel access.

LTE-U enables coexistence with WiFi by means of duty cycling (Fig.~\ref{fig:lte-u-overview}) rather than LBT. Qualcomm~\cite{lteu_qualcom2016} recommends that LTE-U should use period of 40, 80 or 160\,ms, and limits maximal duty cycle to 50\%.
The LTE-U BSs actively observe the wireless channel to estimate its utilization. This estimate is used for dynamic channel selection and adaptive duty cycling. In principle, the least occupied channel is preferred over others.
The mechanism called carrier sense adaptive transmission (CSAT) is used to adapt the duty cycle, by modifying the $T_{\mathrm{ON}}$ and $T_{\mathrm{OFF}}$ values, to achieve fair sharing.

\iftoggle{techreport}{
\begin{wrapfigure}[6]{R}{0.5\textwidth}
    \vspace{-10pt}
	\centering
    \includegraphics[width=1\linewidth]{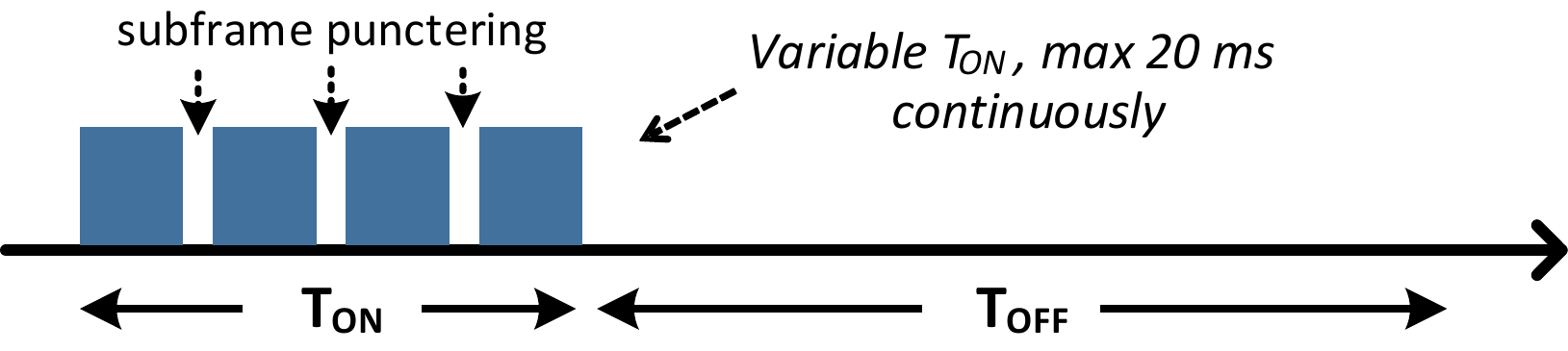}
    \vspace{-20pt}
    \caption{LTE-U adaptive duty cycle (CSAT).}
    \label{fig:lte-u-overview}
\end{wrapfigure}
}%
{
\begin{figure}[h!]
	\centering
	\includegraphics[width=1\linewidth]{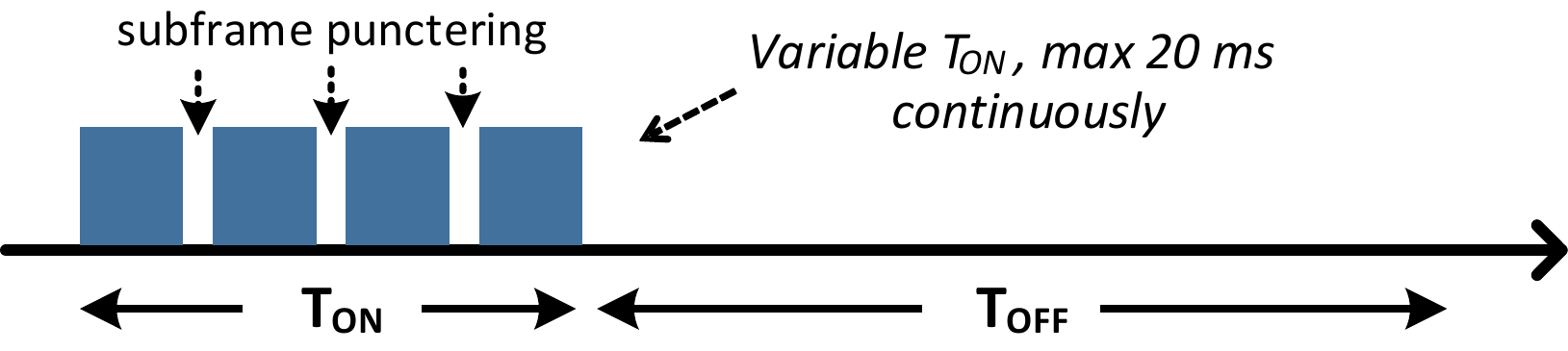}
	\vspace{-20pt}
	\caption{LTE-U adaptive duty cycle (CSAT).}
	\label{fig:lte-u-overview}
	\vspace{-5pt}
\end{figure}	
}

In addition, LTE-U transmissions contain frequent gaps during the \textit{on-phase} (so called subframe puncturing), which allow WiFi to transmit delay-sensitive data. At least 2\,ms puncturing has to be applied every 20\,ms according to Qualcomm's proposal.~\cite{lteu_qualcom2016}

%
%
\subsection{WiFi}

In contrast to LTE-U which uses scheduled channel access WiFi stations perform random channel access using a Listen-Before-Talk (LBT) scheme (i.e. modified CSMA). While coexistence among multiple WiFi sets makes use of both virtual and physical carrier sensing, collisions with other technologies (here LTE-U) can be avoided by the energy-based carrier sensing (CS) known to be less sensitive as compared to preamble-based CS methods.
The periodic (scheduled) LTE-U transmissions may impact the WiFi communication in two following ways: \textit{i)} block medium access by triggering the Energy Detection (ED) physical CS mechanism of WiFi (less available airtime for WiFi due to contention); \textit{ii)} corrupt packets due to co-channel interference (wasted airtime due to packet loss, i.e. a form of inter-technology hidden node). The occurrence of the first or the second effect depends on the received LTE-U signal strength at the WiFi transmitter and receiver respectively.


\section{Problem Statement}

Our target system is shown in Fig.~\ref{fig:system_model}. Here we have WiFi APs being co-located with LTE-U BSs operating in the same unlicensed radio spectrum. Both technologies are serving multiple end-user terminals (not shown in the figure). We incorporate multiple co-located LTE-U cells, even from different operators, as long as they have (time) aligned duty-cycles like proposed by Cano et al.~\cite{cano2015}.

\iftoggle{techreport}{
\begin{wrapfigure}{R}{0.5\textwidth}
    \vspace{-10pt}
	\centering
	\includegraphics[width=0.8\linewidth]{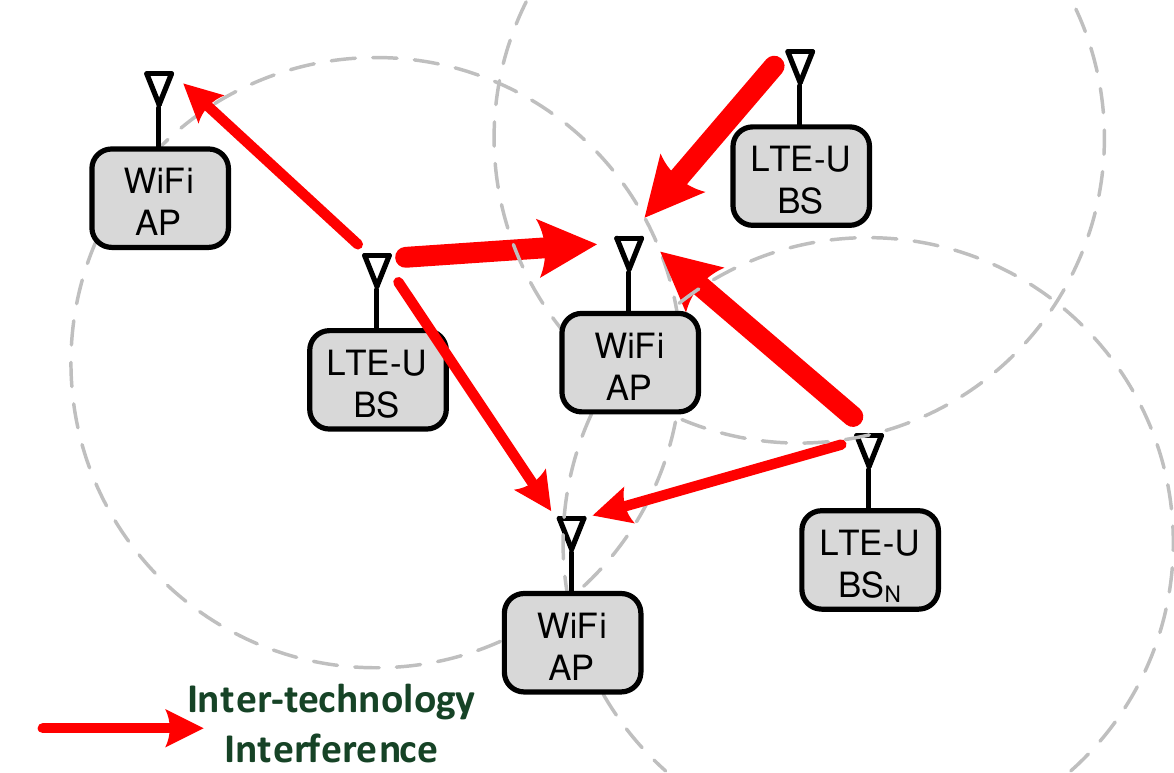}
	\vspace{-5pt}
	\caption{System model --- WiFi APs surrounded by multiple LTE-U cells. The LTE-U DL traffic creates interference on WiFi DL and UL traffic.}
	\label{fig:system_model}
\end{wrapfigure}
}%
{
\begin{figure}[h!]
	\centering
	\includegraphics[width=0.8\linewidth]{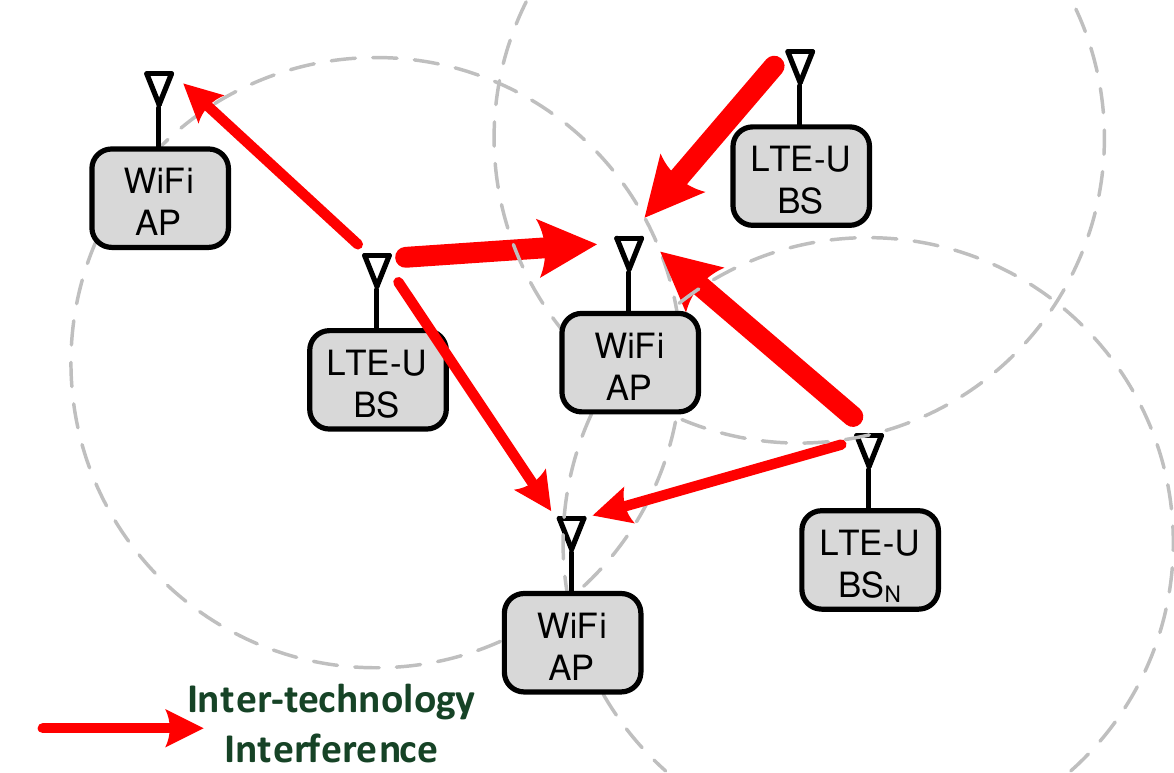}
	\vspace{-5pt}
	\caption{System model --- WiFi APs surrounded by multiple LTE-U cells. The LTE-U DL traffic creates interference on WiFi DL and UL traffic.}
	\label{fig:system_model}
\end{figure}	
}

The main goal of this work is to enable cooperation between co-located LTE-U BSs and WiFi APs operating in the same 5\,GHz band and being owned by different authorities. 

In order to achieve the envisioned cross-technology cooperation, first we need to setup \textbf{common management plane between heterogeneous technologies} (LTE and WiFi), and second, we have to identify the pair of network nodes suffering from performance degradation due to mutual interference, i.e. \textbf{cross-technology proximity detection}. Having achieved those two sub-goals, we will be able to implement advanced cross-technology radio resource and interference management schemes as described in Section~\ref{applications}.


\section{Design Principles}

This section gives an overview of LtFi. First, we present the general architecture of our LtFi system. Then in the following sections, we give a detailed description of its components.

The main design goal of LtFi is transparency, meaning that it should not disturb the operation of higher protocol layers (WiFi as well as LTE-U). Moreover, it should not introduce any additional overhead for the CTC (i.e. over-the-air transmission of additional control frames or signals), but rather use side-channel information which can be used to encode CTC data on top of regular LTE-U frames.

LtFi consists of two parts, namely the LtFi Air-Interface and the LtFi-X2-Interface. The first is used for over-the-air transmission of configuration data from LTE-U BSs to co-located WiFi nodes and for proximity detection, whereas the second is used to create a bi-directional control channel between WiFi nodes and the corresponding LtFi management unit (LtFiMU) over the Internet for the purpose of performing cross-technology radio resource and interference management operations.

\iftoggle{techreport}{
\begin{wrapfigure}{R}{0.5\textwidth}
	\centering
	\vspace{-10pt}
	\includegraphics[width=0.8\linewidth]{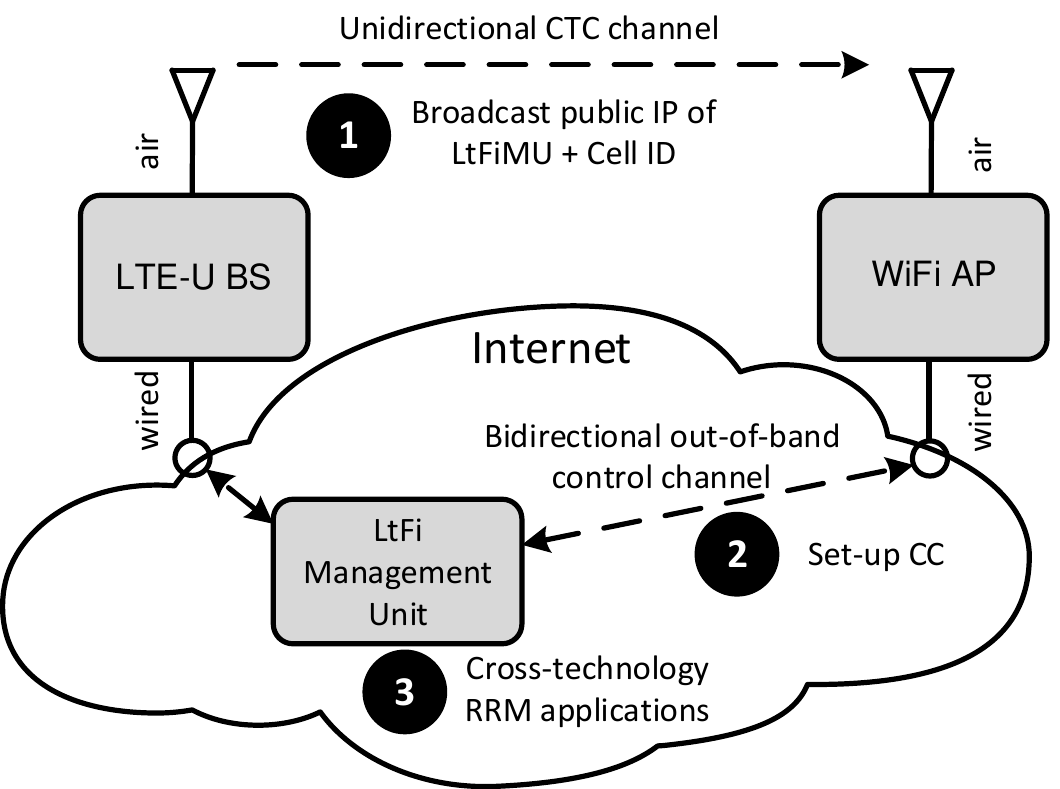}
	\vspace{-5pt}
	\caption{Overview of the system architecture of LtFi --- using the air interface a LTE-U BS transmits configuration parameters which are used to set-up an out-of-band control channels with the corresponding LtFiMU over the Internet.}
	\label{fig:arch0}
	\vspace{-20pt}
\end{wrapfigure}
}%
{
\begin{figure}[!ht]
	\centering
	\vspace{-10pt}
	\includegraphics[width=0.8\linewidth]{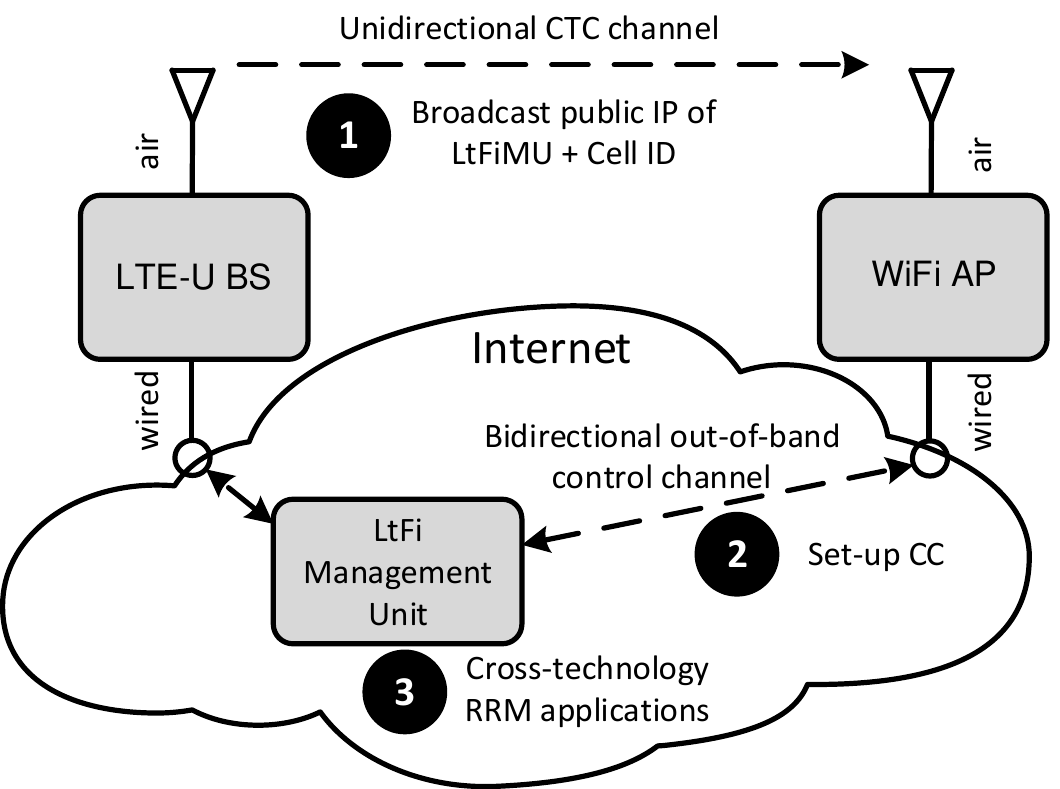}
	\vspace{-5pt}
	\caption{Overview of the system architecture of LtFi --- using the air interface a LTE-U BS transmits configuration parameters which are used to set-up an out-of-band control channels with the corresponding LtFiMU over the Internet.}
	\label{fig:arch0}
	\vspace{-10pt}
\end{figure}
}

Fig.~\ref{fig:arch0} gives a more detailed view on the system architecture of LtFi. LtFi programs the LTE-U eNBs to transmit LtFi configuration parameters (e.g. global IP of LtFiMU) and data used for proximity detection over their air interface. The message is received by co-located WiFi APs which are setting up a bi-directional P2P out-of-band control channels towards the LtFiMU (identified by received global IP) over the Internet. The LtFiMU component acts as a proxy towards the LTE-U network and is responsible for the configuration and management of the individual LTE-U eNBs.

%
%
\section{LtFi -- Air Interface} \label{airInterface}

The LtFi air-interface enables a unidirectional (broadcast) over-the-air communication from LTE-U BS (sender) to WiFi nodes (receiver). Fig.~\ref{fig:architecture} shows how LtFi is integrated into LTE and WiFi systems respectively. The white boxes represent the layers and entities present in existing standards, while the gray boxes are elements of LtFi for which a detailed description is given in the following subsections. Note that, as LtFi is only an add-on to existing standards, it can be easily integrated with already deployed devices by performing software update, i.e. no protocol changes to either LTE or WiFi are needed.

The LtFi air-interface exploits the degree of freedom to place the subframe puncturing into LTE-U, whose relative position is used to modulate information on the CTC. More specifically we place additional puncturing into LTE-U's on-time. The puncturing has a length of 1\,ms, hence the additional delay experienced by LTE-U data packets is negligible. In order to achieve this, the LtFi transmitter is interfaced with the LTE-U scheduler, which is responsible for managing available wireless resources, i.e. Resource Blocks (RBs). The LtFi transmitter is sending a vector $\overrightarrow{s}=[s_1, s_2, ..., s_k]$ of CTC symbols to the eNb scheduler. A CTC symbol $s_i$ represents the relative puncturing position. The eNb scheduler takes $\overrightarrow{s}$ into its radio resource scheduling decision on the unlicensed band, i.e. it stalls (puncture) its transmission for 1\,ms at the time points given in the CTC symbols. Moreover, the interface between LtFi Tx and eNb scheduler is used to negotiate LtFi symbol durations, number of punctures per symbol as well as the configuration of the length of a puncture. This is needed in order to adapt to changing LTE-U traffic load (detailed description is given in Sec.~\ref{trafficAdaptation}) and/or wireless channel conditions.

\iftoggle{techreport}{
\begin{wrapfigure}{R}{0.5\textwidth}
	\centering
	\includegraphics[width=1.0\linewidth]{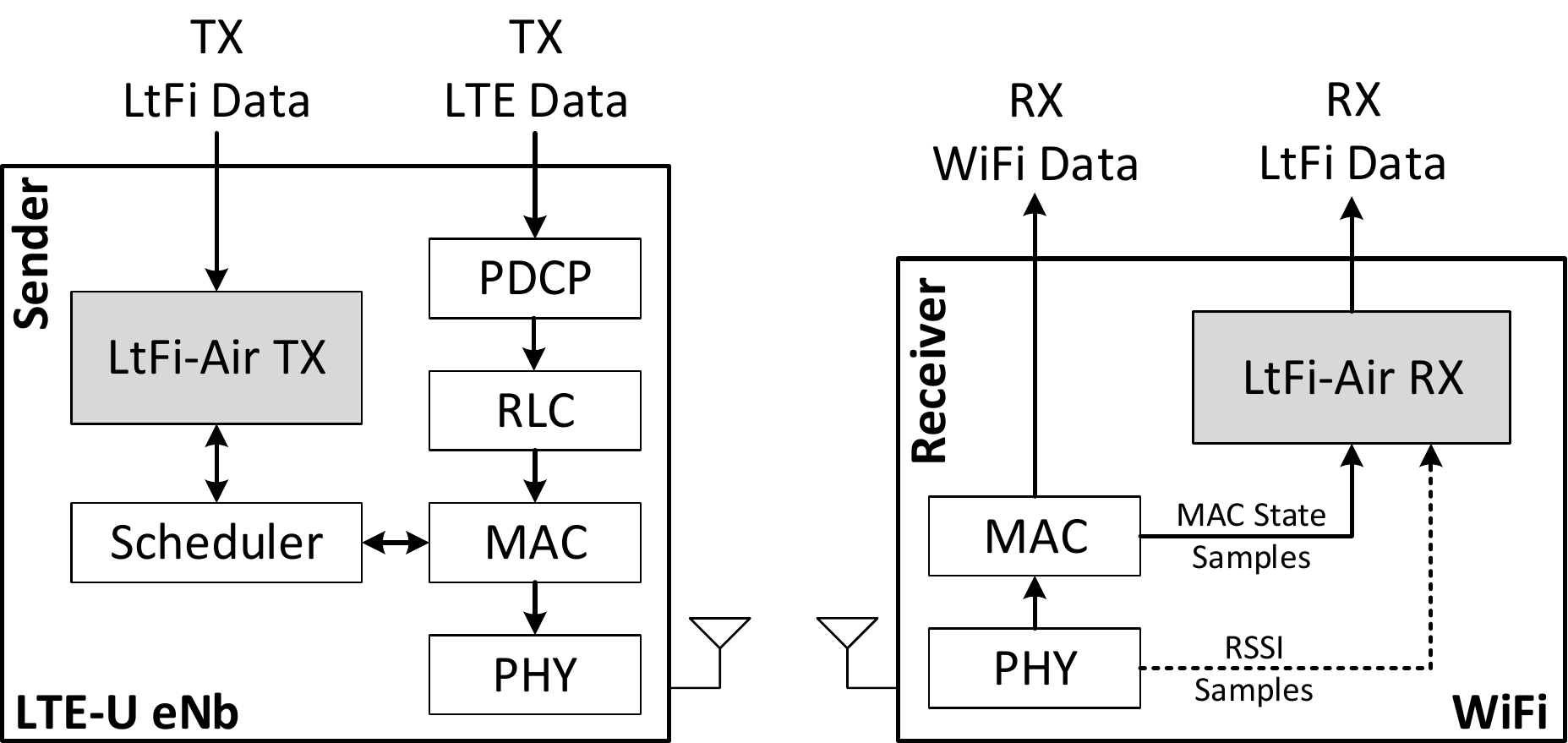}
	\caption{Integration of LtFi into LTE-U eNB and WiFi AP --- LtFi requires an interface to the LTE-U scheduler as well as access to PHY/MAC layer information on the WiFi AP side.}
	\label{fig:architecture}
	\vspace{-10pt}
\end{wrapfigure}
}%
{
\begin{figure}[h!]
	\centering
	\includegraphics[width=1.0\linewidth]{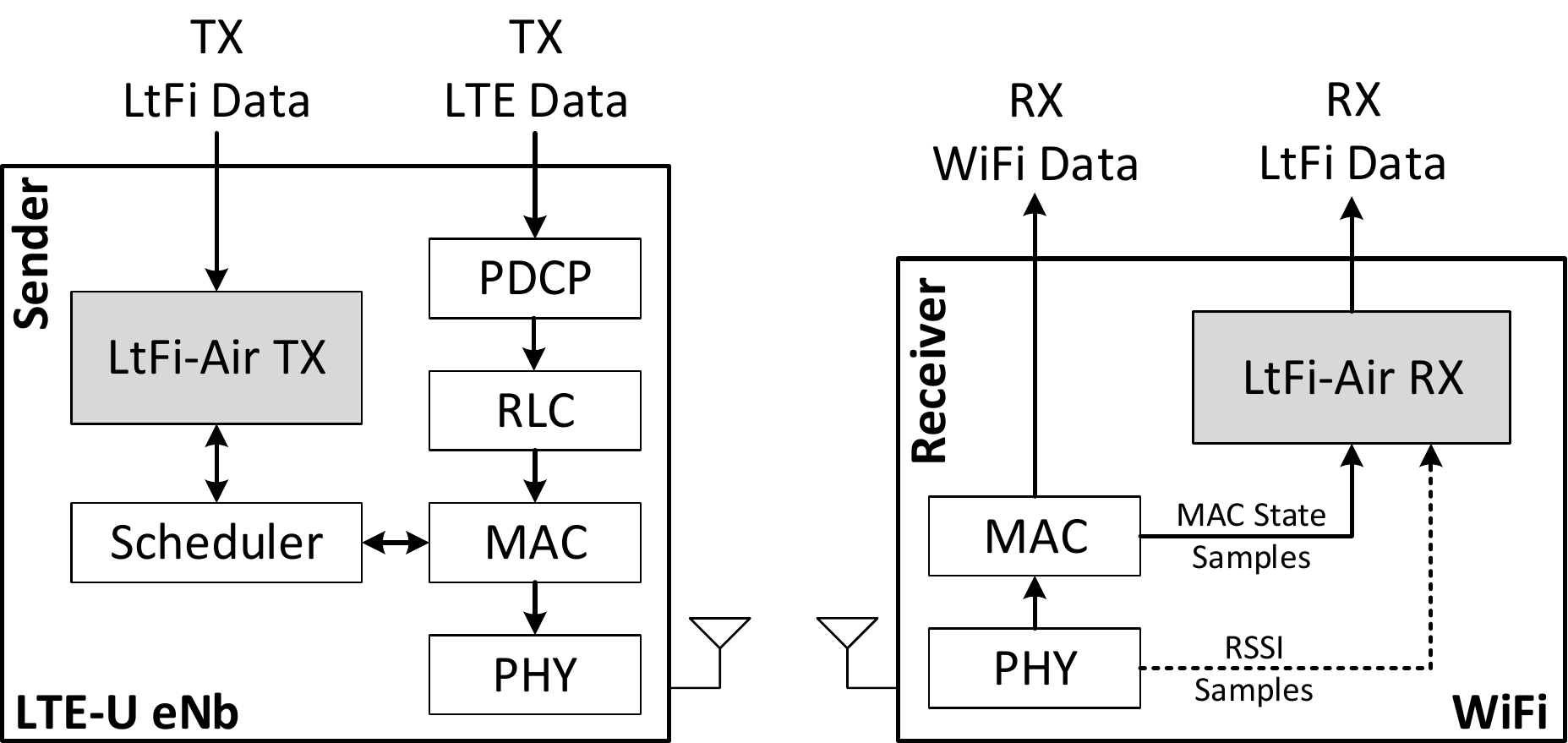}
	\caption{Integration of LtFi into LTE-U eNB and WiFi AP --- LtFi requires an interface to the LTE-U scheduler as well as access to PHY/MAC layer information on the WiFi AP side.}
	\label{fig:architecture}
	\vspace{-5pt}
\end{figure}
}

On the receiver side a WiFi node, typically an AP, needs to decode the LtFi CTC signal. As a direct decoding of the LTE-U frames is not possible due to incompatible physical layers, the LtFi receiver has to detect and decode radio patterns based on receive signal strength (RSSI). To this end, LtFi may rely on spectrum scanning capabilities of WiFi NICs (e.g. Atheros). In this case, however, the receiver has to process large amount of spectral scan data. Instead, LtFi solves this problem by utilizing the possibility to monitor the signal detection logic of modern WiFi NICs (e.g. Atheros 802.11n/ac). More specifically LtFi monitors the relative amount of time the WiFi NIC spent in the \textbf{energy detection (\textit{ED}) without triggering packet reception (\textit{RX}) aka interference (\textit{Intf}) state}, which is entered on reception of a strong non-WiFi signal~\cite{olbrich2017towards}. As LTE-U is so far the only source of interference in the 5\,GHz band, it is safe to assume LTE-U being the non-WiFi signal. Finally, in order to detect the relative position of the puncturing in the LTE-U's on-phase signal the \textit{Intf} state is sampled with sufficient high rate, i.e. sample duration of 0.25\,ms.

%
%
\subsection{TX/RX Chain}

\iftoggle{techreport}{
\begin{wrapfigure}[12]{R}{0.5\textwidth}
	\centering
	\includegraphics[width=0.8\linewidth]{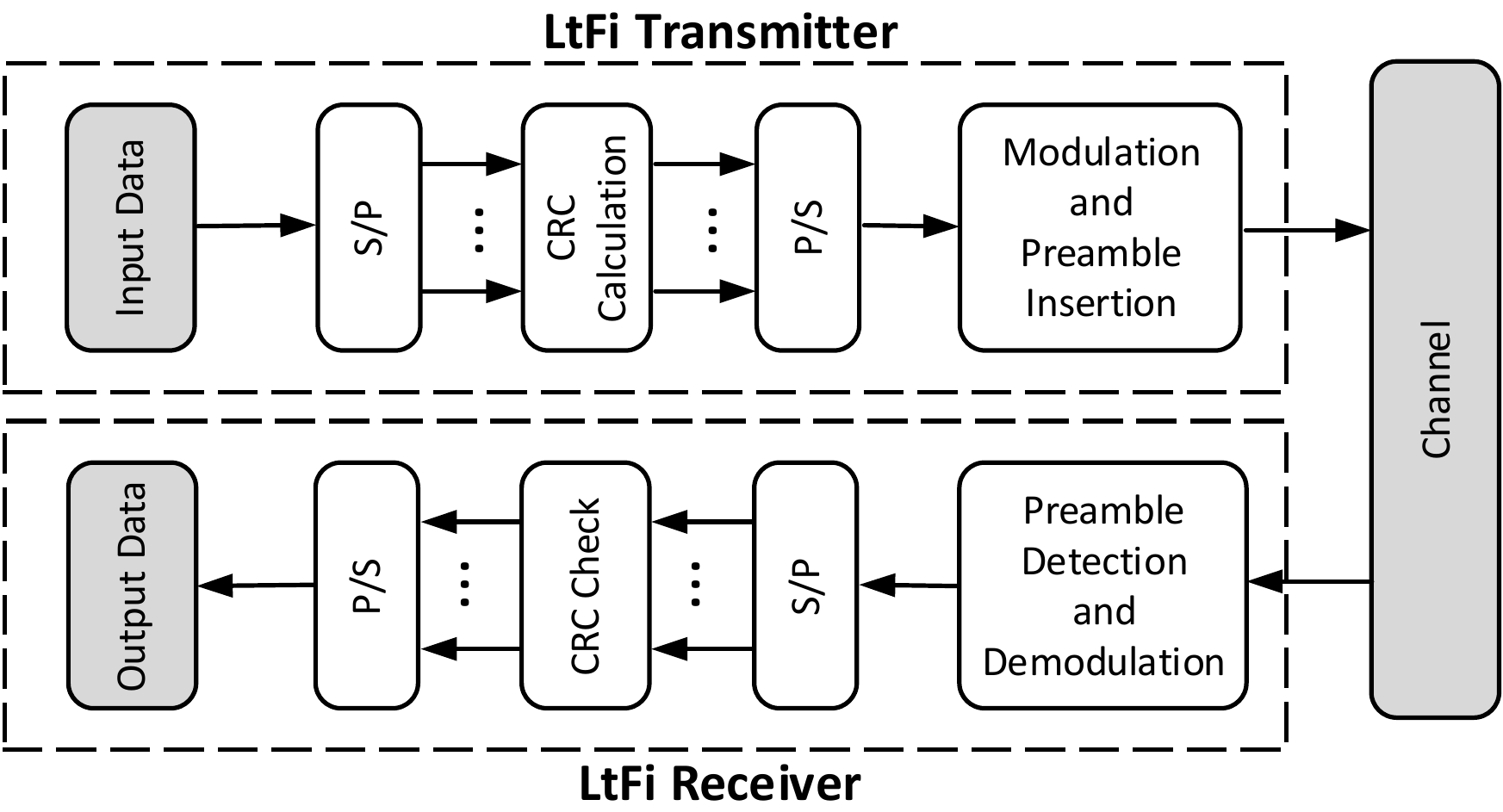}
	\vspace{-10pt}
	\caption{The full TX/RX chain of the LtFi air interface}
	\label{fig:ctc_tx_rx_chain}
	\vspace{-5pt}
\end{wrapfigure}
}%
{
}
The full transmit (TX) and receive (RX) chain of LtFi air interface is shown in Fig.~\ref{fig:ctc_tx_rx_chain}.
Beside the already mentioned modulator and demodulator we have blocks for preamble detection (synchronization) and cyclic redundancy check (CRC-16). Note error detection is needed in order to provide reliable communication over noisy channel, i.e. LTE-U BS being far away but still in interference range to the WiFi node. The preamble is inserted after modulation and is used to mark the start of the frame. The receiver detects preamble using cross-correlation technique.  
The most important blocks are discussed in the next sections.

\iftoggle{techreport}{
}%
{
\begin{figure}[ht!]
	\centering
	\includegraphics[width=0.8\linewidth]{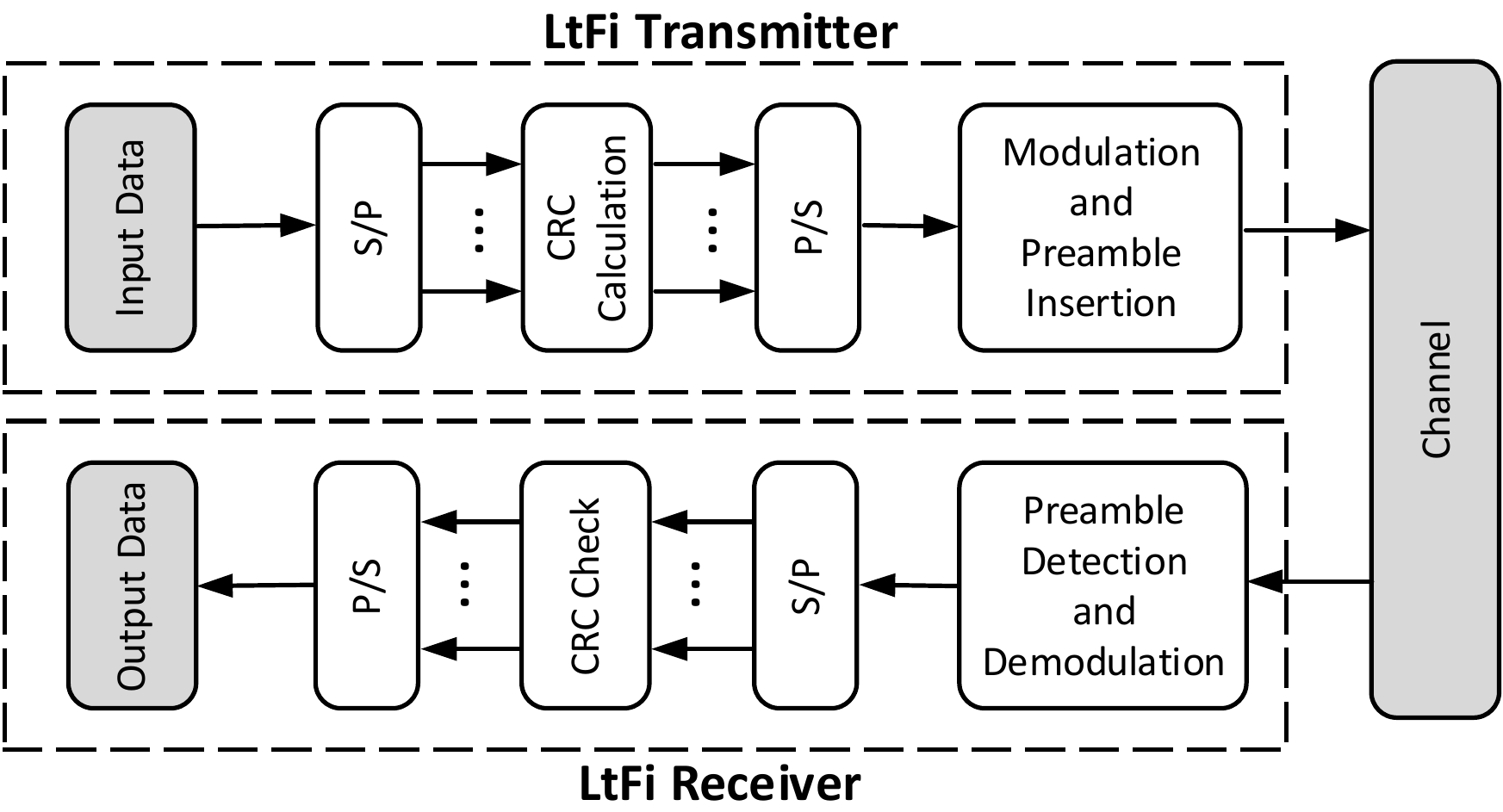}
	\vspace{-10pt}
	\caption{The full TX/RX chain of the LtFi air interface}
	\label{fig:ctc_tx_rx_chain}
	\vspace{-5pt}
\end{figure}
}

%
%
\subsection{Modulation}

This section describes the basics of LtFi's modulation techniques. 
As already stated in Section~\ref{lte_primer}, according to recommendations, a LTE-U BS can transmit continuously only up to 20\,ms and then it has to stop its ongoing transmission for 2\,ms to allow Wi-Fi nodes to send low latency data (e.g. VoIP). Since one puncture every 20\,ms is mandatory, we exploit it to slice the LTE-U signal into chunks of 20\,ms duration that serve as single LtFi symbols.

For the sake of clarity of presentation, we here describe the modulation process assuming that the LTE-U BS has buffered just enough data to fill exactly a single LtFi symbol and we use only a single puncture for encoding CTC data. Moreover, for the following we further assume that there is no other signal transmitted, i.e. no interference.
We will address the more advanced features like the possibility to transmit multiple CTC symbols during a single LTE-U cycle, the usage of more than one puncturing per CTC symbol as well as issues of interference in following subsections.

\iftoggle{techreport}{
\begin{wrapfigure}{R}{0.5\textwidth}
	\centering
	\includegraphics[width=0.95\linewidth]{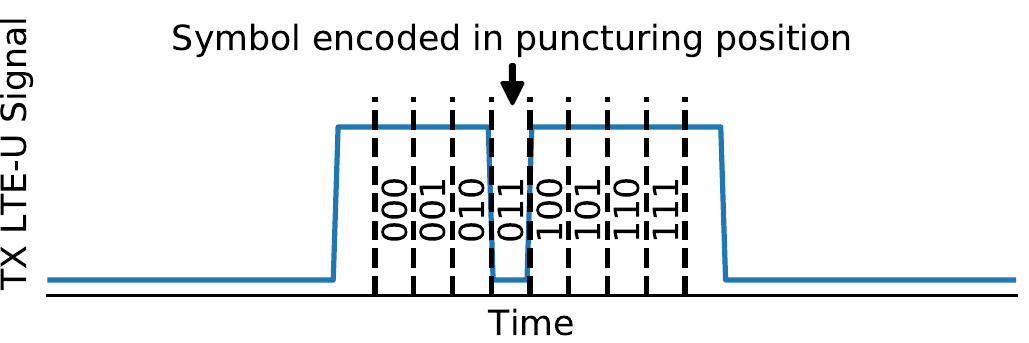}
	\vspace{-10pt}
	\caption{Illustrative example showing modulation at LTE-U side. Here a CTC symbol encodes three bits.}
	\label{fig:modulation}
	\vspace{-20pt}
\end{wrapfigure}

}%
{
\begin{figure}[ht!]
	\centering
	\includegraphics[width=0.95\linewidth]{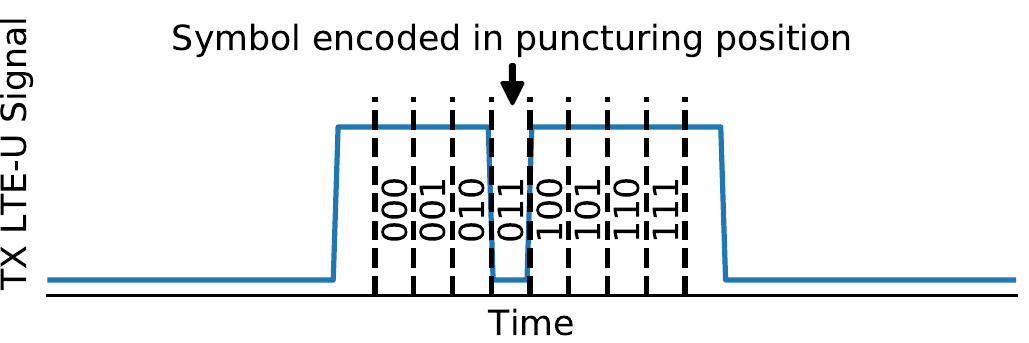}
	\vspace{-10pt}
	\caption{Illustrative example showing modulation at LTE-U side. Here a CTC symbol encodes three bits.}
	\label{fig:modulation}
	\vspace{-5pt}
\end{figure}
}

Without loss of generality, in Fig.~\ref{fig:modulation}, we present a single LtFi symbol with 20\,ms duration that consists of 18\,ms of LTE-U's transmission and one puncture of 2\,ms. With those values, there are ten different possible puncturing positions, but as the receiver has to correctly discover start and end of LtFi symbol, it is not possible to puncture at the first and the last position. Hence, there are only 8 possible positions, what allows for encoding of three bits in one symbol.

%
%
\subsection{Synchronization \& Demodulation}\label{demodulation}

The input to our LtFi receiver is a signal created based on the observation of the relative amount of time the WiFi-NIC MAC spent in specific states, namely, \textit{i)} idle, \textit{ii)} receive (RX), \textit{iii)} transmit (TX) and \textit{iv)} interference (Intf) (Fig.~\ref{fig:mac_states}). 
Listing~\ref{proximity_algo} shows the algorithm used for synchronization and demodulation. Specifically, we take the \textit{Intf} state as it is the time duration the WiFi NIC spents in the energy detection (ED) without triggering packet reception (RX), which is entered on reception of a strong non-WiFi signal like LTE-U. Fig.~\ref{fig:demodulation} shows an example of received \textit{Intf} signal in clean channel. Unfortunately, in practice the \textit{Intf} signal is noisy and needs to be cleaned for which we need the other three states as well (line 11-12). The preamble detector is based on calculating the cross correlation (line 15). After a preamble is detected the receiver is synchronized and starts decoding the symbols. Therefore it computes the cross correlation to each valid symbol and takes the one with highest value. Each symbol is afterwards de-mapped to bits (Fig.~\ref{fig:demodulation}). The receiver continues until it decodes all symbols of the fix length LtFi frame. Note, that the algorithm needs to know the LTE-U configuration parameters like its cycle length which need to be configured manually or can be obtained automatically using WiPLUS~\cite{olbrich2017towards}.

\iftoggle{techreport}{
\begin{figure}
    \centering
    \begin{minipage}{.48\textwidth}
      \centering
      \includegraphics[width=1\linewidth]{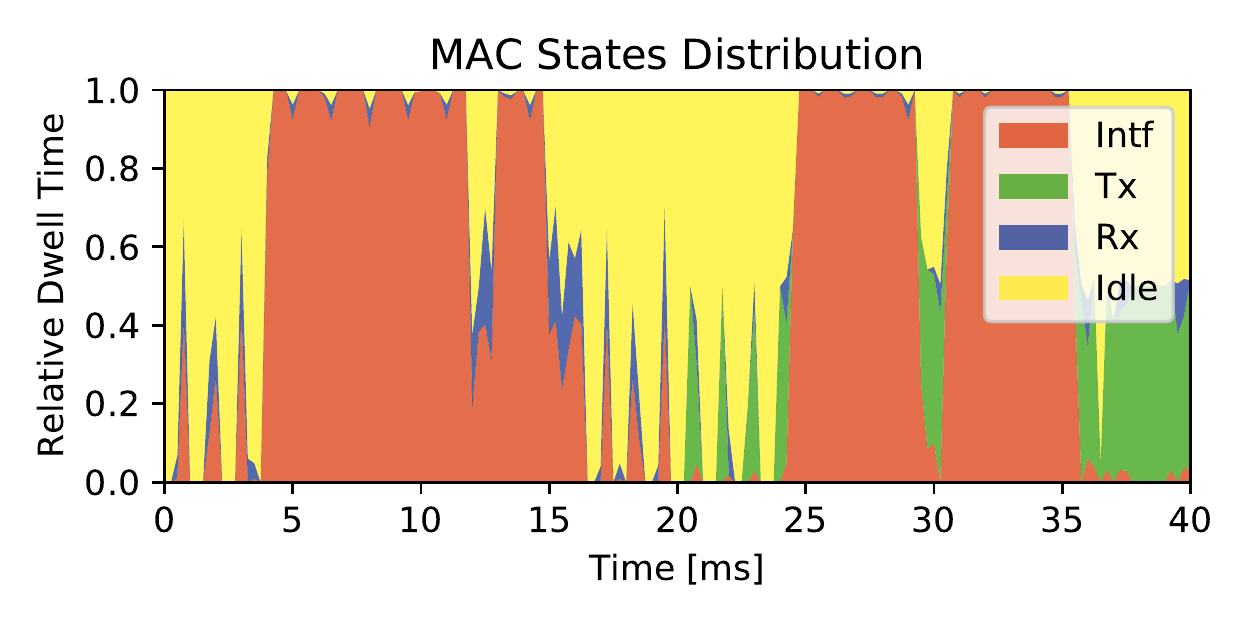}
      \vspace{-20pt}
      \caption{Illustrative example showing the WiFi MAC states distribution.}
      \label{fig:mac_states}
    \end{minipage}%
    \hfill
    \begin{minipage}{.48\textwidth}
      \centering
      \vspace{15pt}
      \includegraphics[width=1\linewidth]{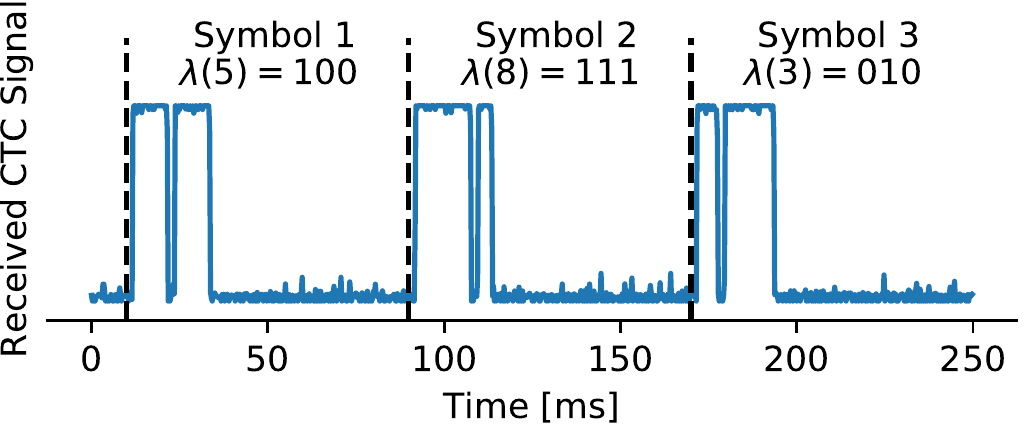}
      \vspace{-20pt}
      \caption{Illustrative example of the received signal and the demodulation at WiFi side.}
      \label{fig:demodulation}
    \end{minipage}
\end{figure}
}%
{
\begin{figure}[h!]
	\centering
	\includegraphics[width=1.0\linewidth]{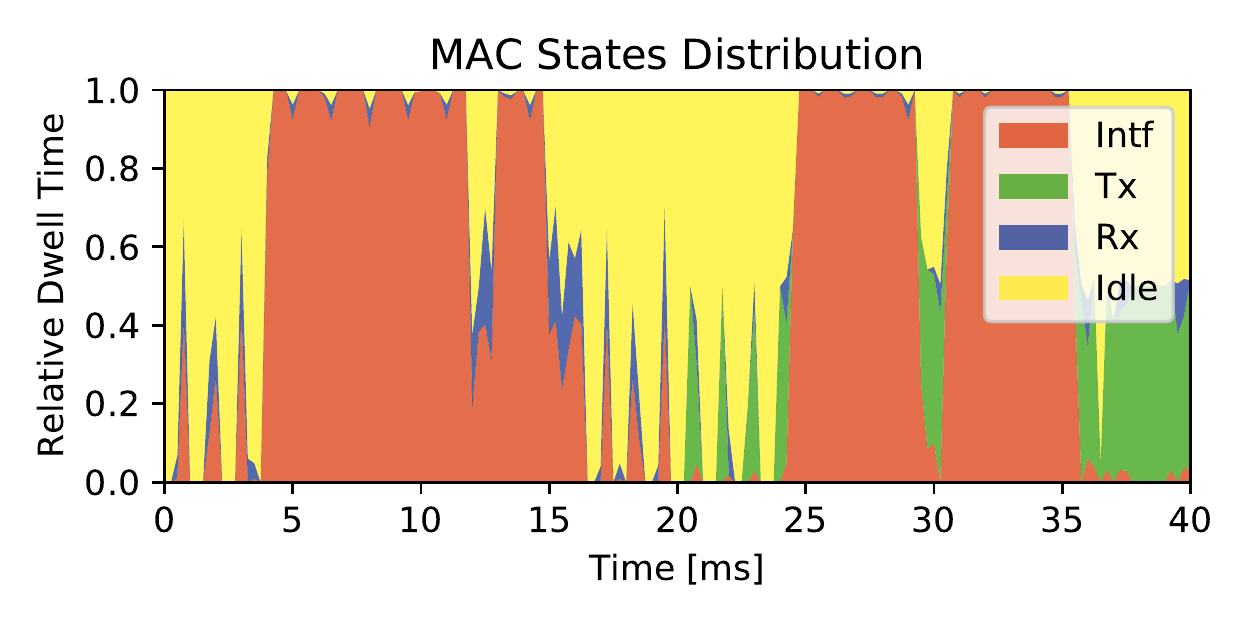}
	\vspace{-25pt}
	\caption{Illustrative example showing the WiFi MAC states distribution.}
	\label{fig:mac_states}
	\vspace{-5pt}
\end{figure}

\begin{figure}[ht!]
	\centering
	\includegraphics[width=0.95\linewidth]{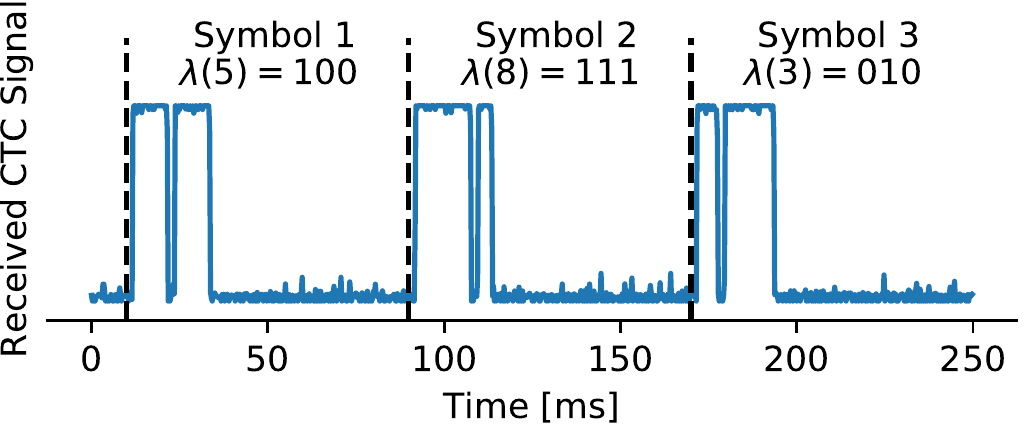}
	\vspace{-10pt}
	\caption{Illustrative example of the received signal and the demodulation at WiFi side.}
	\label{fig:demodulation}
	\vspace{-5pt}
\end{figure}
}

\begin{algorithm}[t]
\DontPrintSemicolon
\caption{LtFi air-interface receiver (preamble detection and demodulation)}
\label{demodulation_algorithm}
\iftoggle{techreport}{
\small
}%
{
\scriptsize
}
\SetCommentSty{scriptsize}
\KwIn{$T_{c}$ \Comment{LTE-U cycle duration}}
\KwIn{$\Delta t = 250\,\mu s$ \Comment{Sampling interval $\rightarrow f_{s} = 4\,kHz$}}
\KwIn{$ED_{t}, RX_{t}, TX_{t}, IDLE_{t}$ {The amount of time spent in each MAC state during last $\Delta t$}}
\KwIn{$\tau_{1}, \tau_{2}, \tau_{3} \in \left \langle 0,1 \right \rangle$ \Comment{Thresholds for signal cleaning}}
\KwIn{$\tau_{p}$ \Comment{Preamble Detection Threshold}}
\KwIn{${P = \left \{ p_1, \ldots, p_N \right \} }$ \Comment{Preamble Reference Signal}}
\KwIn{${M^1,\ldots,M^k}$ \Comment{Set of $k$ possible LtFi symbols}}
\KwIn{$L$ \Comment{LtFi frame length}}

$W \leftarrow \frac{T_{c}}{\Delta t} $ \Comment{Window Size (i.e. samples in LTE-U cycle)}\\
$N \leftarrow 4W $ \Comment{Preamble Length}\\
$t_0 \leftarrow 0 $ \Comment{LtF Symbol Start Marker}\\
$s \leftarrow 0 $ \Comment{Synchronization Flag}\\
$R \leftarrow 0 $ \Comment{Cross-correlation of last synchronization}\\
$l \leftarrow 0 $ \Comment{Number of decoded symbols}\\
$F \leftarrow \{\} $ \Comment{Decoded bits of frame}\\

\While{True} {
    $t \leftarrow t + 1 $\; \Comment{For each new sample}\\
    $S_{t} \leftarrow Intf_{t} $ \Comment{Interference signal (i.e. LTE-U)}\\
    $S_{t} [S_{t} > \tau_1 ] = 1$;
    $S_{t} [1-S_{t} > \tau_2 ] = 0$\Comment{Signal cleaning}\\
    $S_{t} [RX_{t} > \tau_3 ] = 0$;
    $S_{t} [TX_{t} > \tau_3 ] = 0$;
    $S_{t} [IDLE_{t} > \tau_3 ] = 0$\;
    $S_{t} = S_{t} - 0.5 $ \Comment{Remove DC for better CC properties}\\
    $\tilde{P} \leftarrow S_{t-N}, \ldots, S_{t} $ \Comment{Last N samples of recv. signal}\\
    $r = \langle P,\tilde{P} \rangle = \sum_{i=1}^{N} p_i \times \tilde{p}_i $ \Comment{Preamble Detector}\\

    \If{$r \geq \tau_{p}$ \textbf{and} $ s=0 $} {
    $s \leftarrow 1 $;
    $R  \leftarrow r $;
    $t_0 \leftarrow t $\Comment{Preamble detected $\rightarrow$ synchronization}\\
    }
    \If{$r \geq R$ \textbf{and} $ s = 1 $} {
    $R \leftarrow r $;
    $t_0 \leftarrow t $;
    $l \leftarrow 0 $;
    $F \leftarrow \{\} $\Comment{Re-synchronization with higher CC}\\
    }
    \If{$s = 1$ \textbf{and} $t - t_0 = W$} {
    $l \leftarrow l + 1 $;
    $t_0 \leftarrow t $\;
    $\tilde{M} \leftarrow S_{t-W}, \ldots, S_{t} $ \Comment{Received LtFi symbol}\\
    $\langle \tilde{M}, M^k \rangle = \sum_{i=1}^{W} m_i \times \tilde{m}^k_i $ \Comment{Cross-correlation (CC)}
    $k^* = \underset{k}{argmax} (\langle \tilde{M}, M^k \rangle)$ \Comment{Symbol with highest CC}\\
    $B = map(M^{k^*})$ \Comment{Symbol-to-bit mapping}\\
    $F \leftarrow \{F, B\} $\Comment{Append bits to frame}\\
    \If{$l = L$} {
        \textbf{yield} $F$\;
        $l \leftarrow 0 $;
        $s \leftarrow 0 $;
        $F \leftarrow \{\} $\;
        }
    }
}
\end{algorithm}

%
%
%
\subsection{Load-Aware Adaptive Coding Scheme} \label{trafficAdaptation}

So far we assumed that there is enough LTE-U data to be transmitted so that one LtFi symbol of 20\,ms duration can be transmitted during a single LTE-U cycle. In practice, however, as network traffic is bursty (e.g. adaptive video streaming) the duration of the LTE-U \textit{on-phase} can be expected to be variable.

In order to deal with the issue of variable duty cycles, in LtFi we have introduced Load-Aware Adaptive Coding Scheme that selects the proper symbol length and number of punctures depending on the network load in the LTE-U network. The different configurations allow for encoding various number bits in single symbol, hence changing LtFi throughput -- see Section~\ref{analytics} for detailed analysis.

\section{LtFi -- X2 Interface}

The LtFi-X2-Interface is an out-of-band control channel between a LTE-U network represented by the LtFiMU and a WiFi node (mostly AP). Here, we use similar nomenclature as in LTE system, where X2-Interface is out-of-band control channel between BSs. The setup of LtFi-X2-Interface is always initiated by a WiFi node (AP), after successfully decoding the information transmitted over the LtFi air-interface. Specifically it is the public IP address of the LtFiMU and ID of the transmitting LTE-U BS. This information is used to establish the cross-technology channel which can be used by various interference and radio resource management applications (Sec.~\ref{applications}). Note, that the X2 channel can be encrypted using standard protocols like TLS.


\section{Multi-Cell Operation}\label{sec:multicell}

So far we discussed the scenario where a WiFi node is co-located with just a single LTE-U BS. However, in a real environment we can expect to have multiple co-located LTE-U BSs as LTE-U is planned in the same way as normal LTE, e.g. regular hexagonal placement of LTE-U BSs. Hence, the objective of this section is to show how LtFi operates in such a multi-cell scenario.

For the following we assume the worst case scenario where all co-located LTE-U BSs are using the same unlicensed spectrum, i.e. the same radio channel in the 5\,GHz band. Moreover the BSs are time synchronized and aligned, meaning they start their LTE-U CSAT cycle at the same point in time. However, we do not demand that all LTE-U BSs have the same network load, i.e. the same LTE-U on-phase duration. Furthermore, the LTE-U BSs are broadcasting different data over the LtFi air-interface. In such a case, a WiFi node would not be able to decode the signal as it fails to detect the puncturing position, i.e. the puncterings of different BSs are no longer time aligned. The decoding would only succeed in case the WiFi node is very close to one of the BSs as only here the SINR of the CTC is sufficient high. Mathematically we can formulate. Let $\mathcal{V}$ be the set of co-located LTE-U BSs operating on the same unlicensed radio channel. The SINR of the CTC signal from BS $v \in \mathcal{V}$ can be computed as:
\begin{align}\label{eq:SINR}
\mathrm{SINR}^v = \frac{P^v_{\mathrm{RX}}}{\sigma^2 + \sum_{w \in \mathcal{V} \wedge w \neq v}{P^{v}_{\mathrm{CCI}}}}
\end{align}
\noindent where $P^v_{\mathrm{RX}}$ is the total transmit power from BS $v$, $P^{v}_{\mathrm{CCI}}$ is the co-channel interference power level (from the remaining BSs) and $\sigma^2$ is the AWGN power level.

In order to be able to decode the LtFi-CTC signal transmitted by $v$ the $\mathrm{SINR}^v$ must be sufficient high, i.e. WiFi node has to be located very close to $v$. At the cell-edge, where signals from neighboring cells interfere, the SINR is low, thus the signal cannot be decoded. Consider the following illustrative example network given in Fig.~\ref{fig:ctc_ex2} where a WiFi node A is surrounded by three LTE-U BSs. The SINR is very low and hence the decoding would fail as the distance to any of the three BSs is the same. 

With this in mind there are several options. One option would be to let all LTE-U BSs transmit the same information over the CTC so that co-channel interference can be completely avoided. However, with such approach we would lose the proximity detection capability, i.e. the WiFi node would like to know the identities of the BSs in its proximity, i.e. interference range. Another option, it to separate the CTC signal from different BSs in time domain (TDMA), i.e. while one BS is transmitting its CTC signal the other have to remain quit. Only BSs that are far away from each other can reuse time slots as the interference between them is negligible (spatial reuse). Unfortunately, this is inefficient especially in high density deployments as also no LTE data communication can be transmitted during the quit periods in the unlicensed band. 

\iftoggle{techreport}{
\begin{figure*}[!ht]
	\centering
	\includegraphics[width=0.95\linewidth]{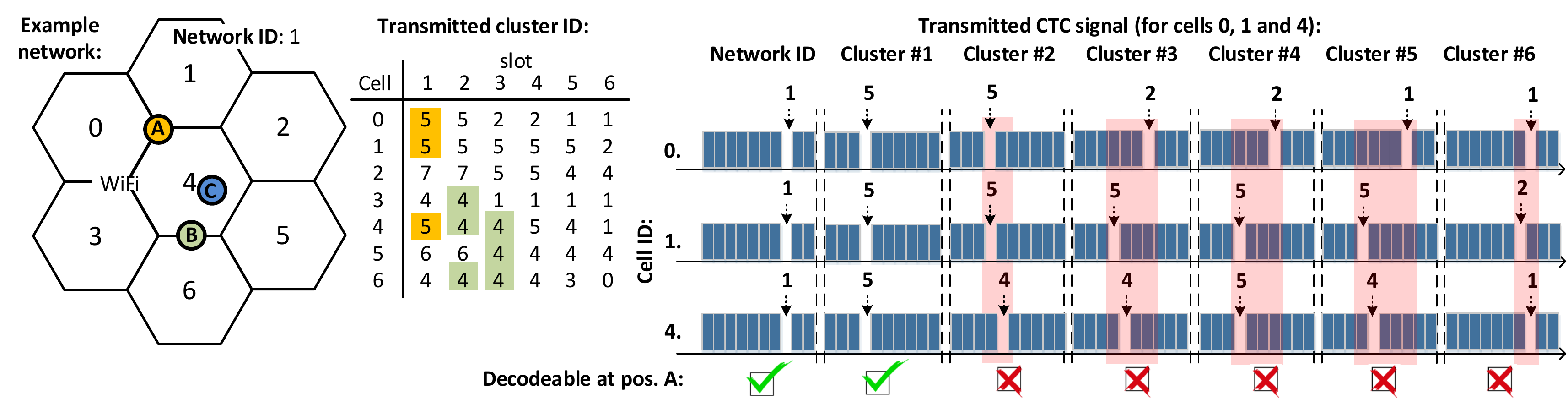}
	\vspace{-15pt}
	\caption{Example network consisting of multiple co-located LTE-U BSs and WiFi node placed at the three different locations (A,B,C). For illustration each CTC symbol encodes 3 bits of information. Table shows for each cell (BS) ID the cluster IDs for the six configurations.}
	\label{fig:ctc_ex2}
\end{figure*}
}%
{
\begin{figure*}[!ht]
	\centering
	\includegraphics[width=0.95\linewidth]{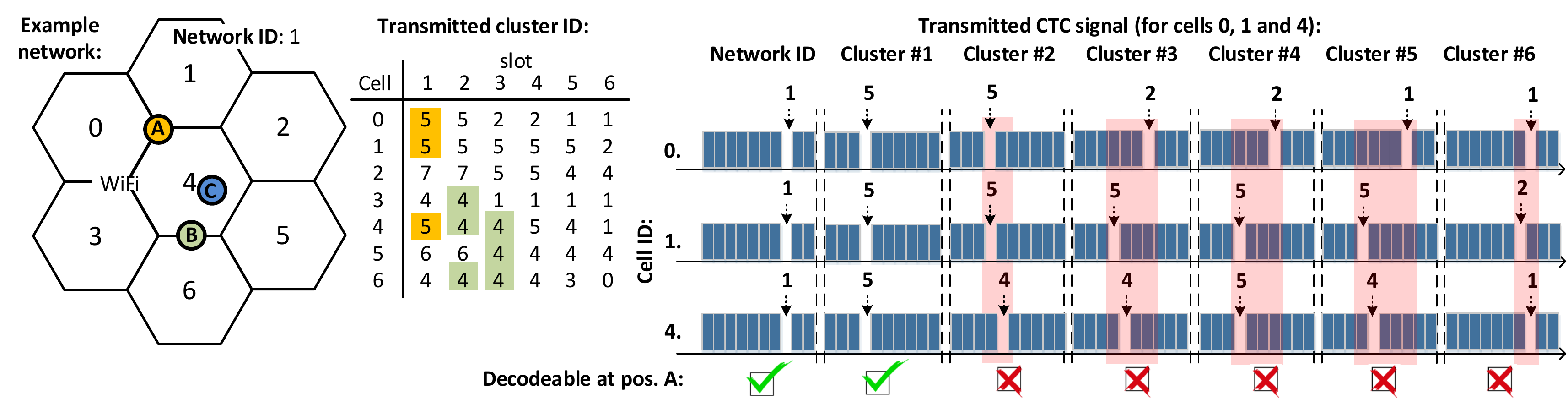}
	\vspace{-5pt}
	\caption{Example network consisting of multiple co-located LTE-U BSs and WiFi node placed at the three different locations (A,B,C). For illustration each CTC symbol encodes 3 bits of information. Table shows for each cell (BS) ID the cluster IDs for the six configurations.}
	\label{fig:ctc_ex2}
	\vspace{-10pt}
\end{figure*}
}

LtFi follows a third approach. It divides the problem of proximity detection into two sub problems: \textit{i)} detection of the LTE-U network identified by the public IP address of its LtFiMU and \textit{ii)} detection of the LTE-U BS IDs in interference range. The first is solved by programming all LTE-BS to transmit the same CTC signal, i.e. public IP address. In absence of any interference this data can be easily decoded by any WiFi node. The second is solved by introducing BS clustering in the LTE-U network where adjacent BSs are grouped in clusters of e.g. size 3. We demand that members of the same cluster have to send the same data over the CTC whereas different data can be send by different clusters. For WiFi nodes located inside those clusters the SINR is improved due to absence of intra-cluster interference. This enables the WiFi node A from Fig.~\ref{fig:ctc_ex2} to decode the data on the CTC channel from the cluster containing cells 0, 1 and 4. Thereby the WiFi node can derive the set of LTE-U cells it is interfering with, here 0, 1 and 4. However, such a static non-overlapping clustering is not sufficient as WiFi nodes located at cluster edges will suffer from inter-cluster interference. LtFi solves inter-cluster interference by using a dynamic (overlapping) clustering, i.e. the members of a given cluster are not fixed but change periodically. For a cluster size of three we have six overlapping cluster configurations with changing members to cover all the six cell-edges. As with overlapping clusters a BS is no longer member of exactly one cluster, the overlapping clusters need to be orthogonalized in time. 

\iftoggle{techreport}{
\begin{wrapfigure}[5]{R}{0.5\textwidth}
	\centering
	\includegraphics[width=1\linewidth]{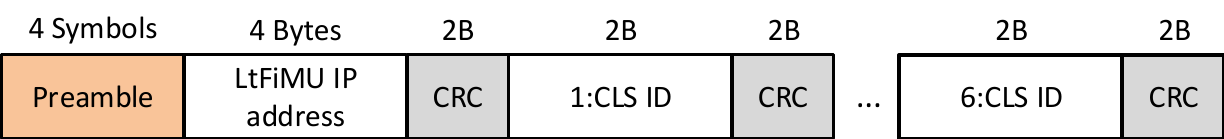}
	\vspace{-20pt}
	\caption{Structure of LtFi Air-Interface frame.}
	\label{fig:ctc_frame}
\end{wrapfigure}
}%
{
\begin{figure}[h]
	\centering
	\includegraphics[width=1\linewidth]{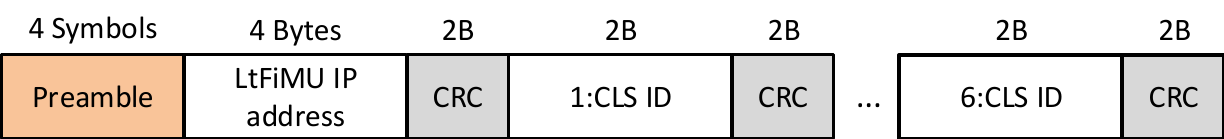}
	\vspace{-20pt}
	\caption{Structure of LtFi Air-Interface frame.}
	\label{fig:ctc_frame}
\end{figure}
}

Fig.~\ref{fig:ctc_frame} shows the framing in LtFi. The frame consists of two parts. The first part contains the network ID (public IP address of LtFiMU) which is used to detect the LTE-U network. As all BSs are transmitting the same information there is no interference on the CTC and the network ID can be decoded by any WiFi node. The second part consists of the six overlapping cluster IDs. Depending on the location of the WiFi node only a subset of the six cluster IDs can be decoded from which the WiFi is able to derive the set of interfering BSs. Note, that the different fields are protected by separate CRC. This is required in order to make sure that each part of the CTC message can be decoded independently as the receiver experiences different SINR for each of them. As the size of entire message are 30\,Bytes plus four symbols for the preamble. Note, that start and end of element of the message has to be aligned with symbol boundaries.

Fig.~\ref{fig:ctc_ex2} illustrates LtFi's dynamic clustering for three different WiFi node locations. At location \textit{A} the WiFi node is able to decode the network ID and just the cluster ID 5. No other cluster IDs can be decoded. Hence the WiFi node assumes to be at the edge between the cells being member of cluster 5 in configuration (slot) 1 which are the cells 0, 1 and 4. At location \textit{B} the WiFi node is at the edge between cells 4 and 6. Here the cluster ID 4 can be decoded in configuration 2 and 3 which corresponds to the clusters (3,4,6) and (4,5,6). Hence it assumes to be at the edge between those two clusters, i.e. edge between BS 1 and 4. Note that in this illustrative example we assume that the signals received from non-adjacent BSs (tier 2) are weak and hence have only minor impact.

%
%
\subsubsection*{Estimating LTE-U BSs in proximity}\label{sec_prox}

A WiFi node continuously decodes the information it receives on the LtFi air-interface. After decoding the network ID it uses the LtFi-X2 interface to create a control channel over the wired Internet to the corresponding LtFi management unit (LtFiMU). In the next step the WiFi node retrieves from the LtFiMU a codebook which is needed to be able to derive the actual LTE-U BS IDs from the $<$configuration/slot number, cluster ID$>$ tuples the WiFi node received over the LtFi air-interface. Note, this is required as we apply dynamic (overlapping) clustering in the LTE-U network, i.e. the membership of BSs to clusters is not static and is different for the six configurations (time slots).

We represent the codebook as a matrix which is constructed as follows. The entry in row $i$ and column $j$ contains the set of cell IDs being member of the cluster $i$ in configuration $j$. Note, for cluster size of three we need six overlapping cluster configuration, hence $j \in (1,6)$. The matrix shown below is the codebook for the example given in Fig.~\ref{fig:ctc_ex2}. Here only the entries for the clusters with ID 4 and 5 are shown:
\begin{equation*}
B=
\setlength\arraycolsep{3pt}
\begin{bmatrix}
\arraycolsep=0pt\def\arraystretch{0pt}
& & & \vdots & & &\cr
\{3,6,\}&\{3,4,6\} & \{4,5,6\} & \{5,6,\} & \{2,4,5\} & \{2,5,\}\cr
\{0,1,4\} & \{0,1,\} & \{1,2,\} & \{1,2,4\} & \{1,,\} & \{6,,\}\cr
& & & \vdots & & &\cr
\end{bmatrix}
\end{equation*}

The final step is the computation of set of BS IDs from the data received over the LtFi air interface for which the algorithm is shown in Listing~\ref{proximity_algo}.

\begin{algorithm}[t]
\DontPrintSemicolon
\caption{Algorithm executed by each WiFi AP to derive the LTE-U cells in proximity.}
\iftoggle{techreport}{
\small
}%
{
\footnotesize
}
\label{proximity_algo}
\KwIn{$C=\{(i_1,c_1), \ldots, (i_n,c_m) \}$ \Comment{ Set of decoded configuration number and cluster IDs}}
\KwIn{$B$ \Comment{Codebook received from LTE-U network}}
$X \leftarrow \{ B(i,c), (i,c) \in C\}$ \Comment{Translate $C$ into set of sets of cells IDs using codebook $B$}\\
$Y \leftarrow \bigcup_{A \in X}A$ \Comment{$Y$ contains cell IDs being member of any element in $X$}\\
\Return $Y$ \Comment{Return the set of cell IDs in proximity}
\end{algorithm}

For our example from Fig.~\ref{fig:ctc_ex2} the WiFi node at location B would have $C=\{(2,4), (3,4)\}$ and with codebook $B$ it would calculate:
\begin{align*}
    X = \{ \{3,4,6\}, \{4,5,6\} \} \\
    Y = \{3,4,5,6\}
\end{align*}
Hence the LTE-U BSs in proximity of WiFi node at location B are $3,4,5$ and $6$.

Note, the knowledge about the actual LTE-U BSs in proximity is a prerequisite for advanced cross-technology interference and radio resource management schemes (cf. Sec.~\ref{applications}).


\section{Prototype Implementation Details}

This section gives a brief overview of the LtFi prototype implementation.

\subsection{LtFi transmitter -- LTE-U BS}

The LTE-U BS waveform together with the LtFi CTC signal was pre-computed offline using Matlab and afterwards radiated over the air using R\&S SMBV100A Vector Signal Generator. The durations of the CSAT cycle and the on phase were fixed and set to 80\,ms and 19\,ms, respectively. Hence, during a single CSAT cycle LtFi was able to transmit a single symbol. The generated LTE-U waveform was transmitted in infinite loop. With such a setup we were able to evaluate the performance of the LtFi air-interface on the link-level.

\subsection{LtFi receiver -- WiFi node}

For the WiFi node we selected WiFi chipsets based on Atheros AR95xx as they allow direct monitoring of the signal detection logic of the WiFi NIC at a very fine granular level. We sampled the Atheros registers with a rate of 4\,kHz and process the data in chunks of 1\,s window sizes. Therefore we migrated the RegMon tool~\cite{Huehn2013} to SMP systems (Ubuntu 16.04) and provided a patch to the upstream ath9k wireless driver. Moreover, we replaced the ring buffer in Regmon by relay file system (relayfs) as it provides an efficient mechanism for transferring large amounts of data from kernel to user space. More details on Atheros signal detection logic can be found in \cite{olbrichwiplus} and the patent from Atheros~\cite{patent-atheros1}. 

The LtFi receiver (see Listing~\ref{demodulation_algorithm}) was implemented entirely in Python language. Our prototypical (not optimized) version of receiver runs in real-time occupying only up to 15\% of single core of i5-4250U (1.30\,GHz) CPU time.

\subsection{LtFi Management Unit}

LtFiMU was implemented using the UniFlex controller framework~\cite{uniflex}. Currently, it only serves connection setup from WiFi nodes and provides it with the codebook for decoding cell (BS) IDs of neighboring LTE-U BSs.


\section{LtFi Applications}\label{applications}

LtFi establishes a cross-technology control channel between co-located LTE-U and WiFi networks. Such a channel can be used to optimize co-existence between the two wireless technologies by means of cross-technology interference and radio resource management. This section gives an overview of possible approaches.

\subsection{Cross-technology Contention \& Interference Management}

Co-located LTE-U and WiFi networks may suffer performance degradation due to either contention, i.e. insufficient free airtime, or co-channel interference, i.e. packet corruption due to the insufficient sensitivity of the energy-based carrier sensing in WiFi, i.e. cross-technology hidden node. Both problems can be solved using LtFi. The former is achieved by adapting the LTE-U CSAT to the actual network load in both the LTE-U and WiFi network to enable a fair use of the shared radio resources. Moreover, the WiFi MAC parameters like CWmin/CWmax and TXOP can be tuned. 

Co-channel interference can be mitigated in two ways. First, by adapting the threshold used for energy-based carrier sensing in the WiFi network. Second, by performing an interference-aware channel assignment to LTE-U and WiFi. Specifically, it is beneficial to put those networks (LTE-U or WiFi) suffering from cross-technology hidden node problem on different ISM radio channels. In a similar way the cross-technology exposed terminal problem can be solved.

Finally, we have found many works in literature (~\cite{wifilteu, 7558177, Mahesh_lte_wifi, 7496918, CU_LTE, 7247524, cano2015, wifi_meets_lte}) that would directly benefit from our LtFi system. For example, Duet~\cite{Duet} assumes that LTE-U BSs are equipped with an additional WiFi interface used to count number of active WiFi stations based on overheard frames. With usage of LtFi the additional interface is superfluous, as the BS can get those data directly from neighboring APs using the CTC.

\subsection{Assuring QoS}

As LTE-U constitutes a new source of interference with strong impact on WiFi ensuring QoS in WiFi is challenging. Especially, we can assume that network traffic requiring low-latency (VoIP, video conferencing, etc.) will suffer the most. Using LtFi a WiFi network can communicate its QoS requirements to the co-located LTE-U network, e.g. in case of low-latency traffic in WiFi network additional puncturing can be added dynamically in the LTE-U ON phase.


\section{Analytics}\label{analytics}

Here we provide a theoretical analysis of the achievable data rate on the LtFi air-interface. As mentioned in Sec.~\ref{airInterface}, there is one mandatory puncture of 2\,ms duration that has to be applied to LTE-U's transmission every 20\,ms. In LtFi we keep this mandatory puncturing so that those 20\,ms chunks represent the LtFi symbols. Inside the symbol we can add up to $k$ additional punctures of 1\,ms duration to encode CTC data bits. By increasing $k$ more bits can be encoded into a single LtFi symbol. The number of available symbols $M$ can be computed as binomial coefficient (\ref{eqn:symNum}), where $n$ is the number of possible puncturing positions (here 18). The modulation rate (bits per symbol) can be computed using equation~(\ref{eqn:bitNum}):
\begin{align}
\label{eqn:symNum}
    M = \binom{n}{k} = \frac{n!}{k!(n-k)} \\
    0 \leq  k  \leq n \\
\label{eqn:bitNum}
    K[bit] = \left \lfloor log_{2}(M) \right \rfloor
\end{align}

Note, that by increasing $k$ while keeping the LTE-U load constant, the LTE-U ON-phase is artificially increased, i.e. the time between start of first and end of last transmission in single cycle. Moreover, that as WiFi uses random channel access it may happen that it starts its transmission just before the start of the ON-phase or within a puncturing, leading to cross-technology interference and possible packet loss. For our analysis we took the worst-case scenario, i.e. each WiFi packet transmission being overlapping with an ongoing LTE-U transmission is assumed to get lost. Moreover, as WiFi frame duration we assumed 384\,$\mu$s\footnote{According to \cite{cmorse} 97\% of the WiFi frames have a duration of less than 384\,$\mu$.}. Thus, in the worst case only around 0.6\,ms out of 1\,ms puncture is available for WiFi transmission. Finally, a collision can also lead to packet loss in LTE-U network. Especially the first slot (0.5\,ms) after a puncturing is prone to collisions for which we suggest to use either a more robust MCS or power bursting.

\iftoggle{techreport}{
\begin{wrapfigure}[15]{R}{0.5\textwidth}
	\centering
	\vspace{-10pt}
	\includegraphics[width=1\linewidth]{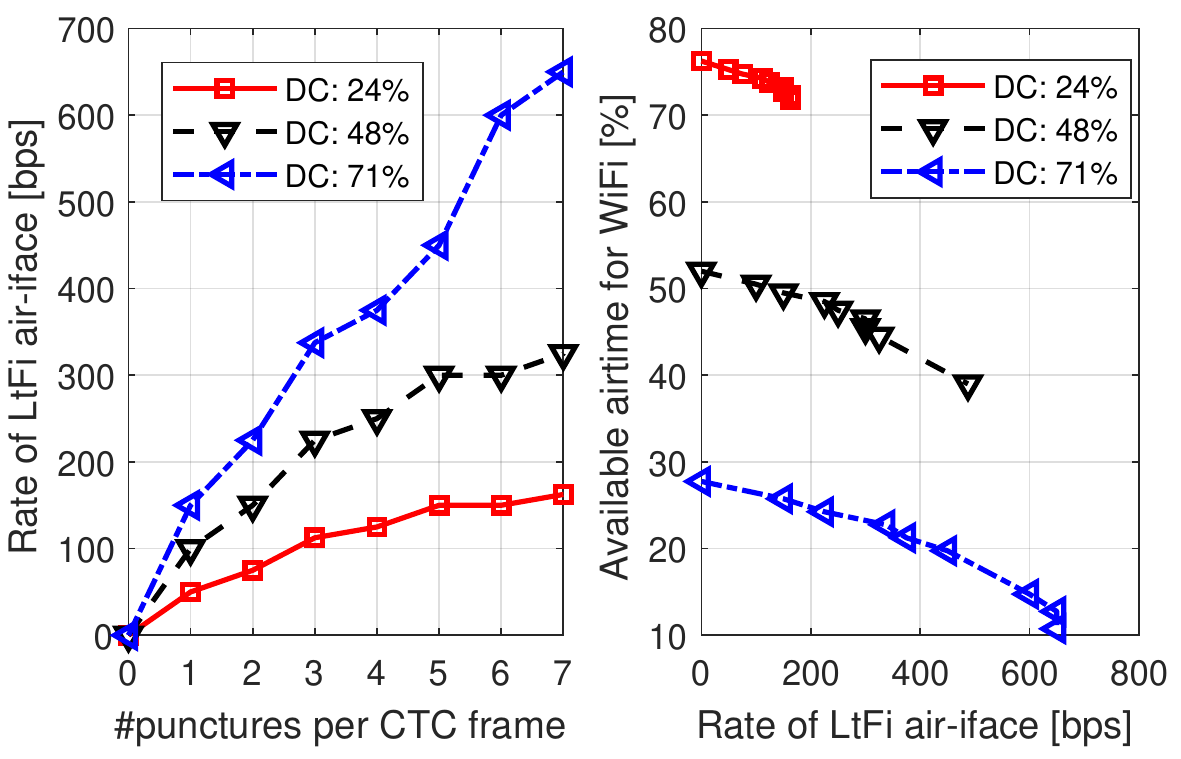}
	\vspace{-20pt}
	\caption{Analytical results of the LtFi air-interface.}
	\label{fig:ctc_analytical_bino}
	\vspace{-10pt}
\end{wrapfigure}
}%
{
\begin{figure}[h]
	\centering
	\includegraphics[width=0.9\linewidth]{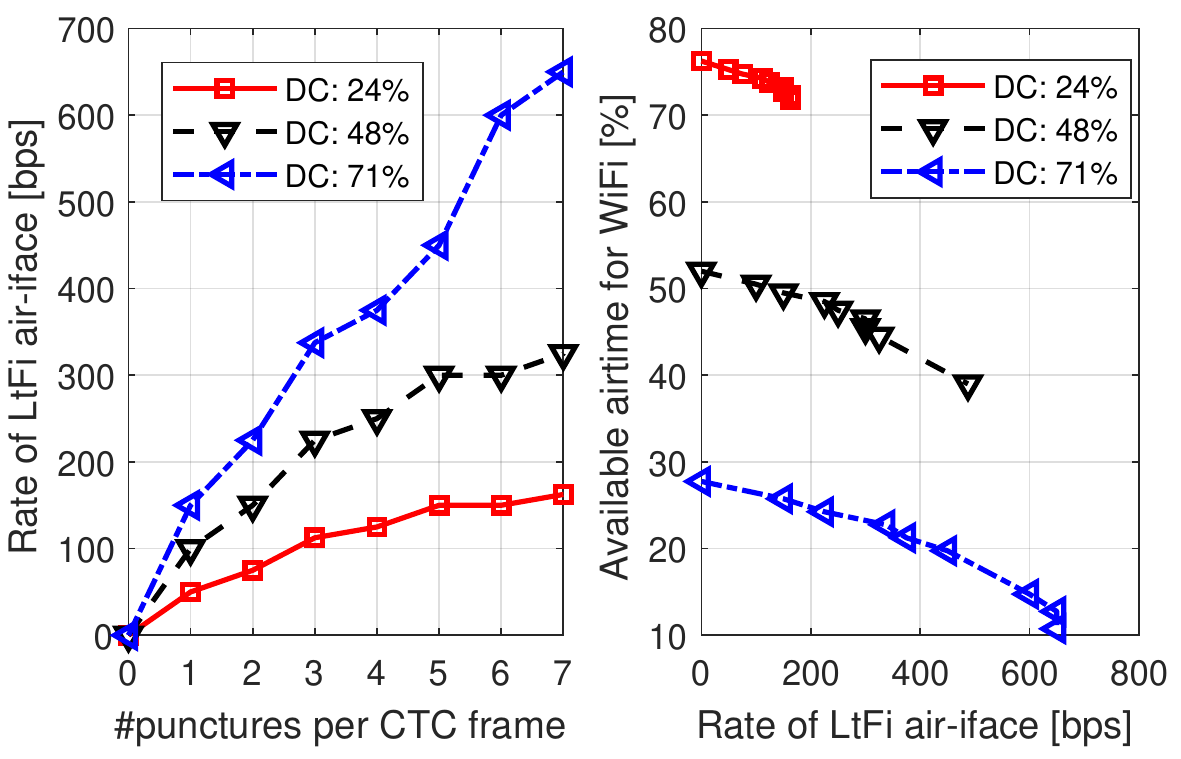}
	\vspace{-10pt}
	\caption{Analytical results of the LtFi air-interface.}
	\label{fig:ctc_analytical_bino}
	\vspace{-5pt}
\end{figure}
}

Fig.~\ref{fig:ctc_analytical_bino} shows the data rate of the LtFi air-interface with increasing number of punctures $k \in \left \{0,1,..,9\right \}$  for different LTE-U duty cycle lengths. We also present the air-time available for WiFi. We can observe that for large $k$ and duty cycle (DC) a data rate of up to 650\,bps can be achieved. For a small duty cycle of 24\% the data rate is between 50 and 160\,bps depending on the number of punctures per symbol $k$.

\medskip
\noindent \textit{\textbf{Takeaways: }} The data rate of the LtFi air-interface is sufficient high to deliver connection and identification information to co-located WiFi nodes, i.e. it takes at most 10\,s with the lowest and less than 1\,s with the highest data rate. We argue that it is enough to support not only static but also nomadic environments (e.g. smartphone in WiFi tethering aka softAP).


\section{Experiments}\label{sec:exp}

LtFi was prototypically implemented and evaluated by means of experiments. This section presents results with the focus on the LtFi air-interface.

\subsection{Methodology}

\iftoggle{techreport}{
\begin{wrapfigure}[12]{R}{0.5\textwidth}
	\centering
	\includegraphics[width=1\linewidth]{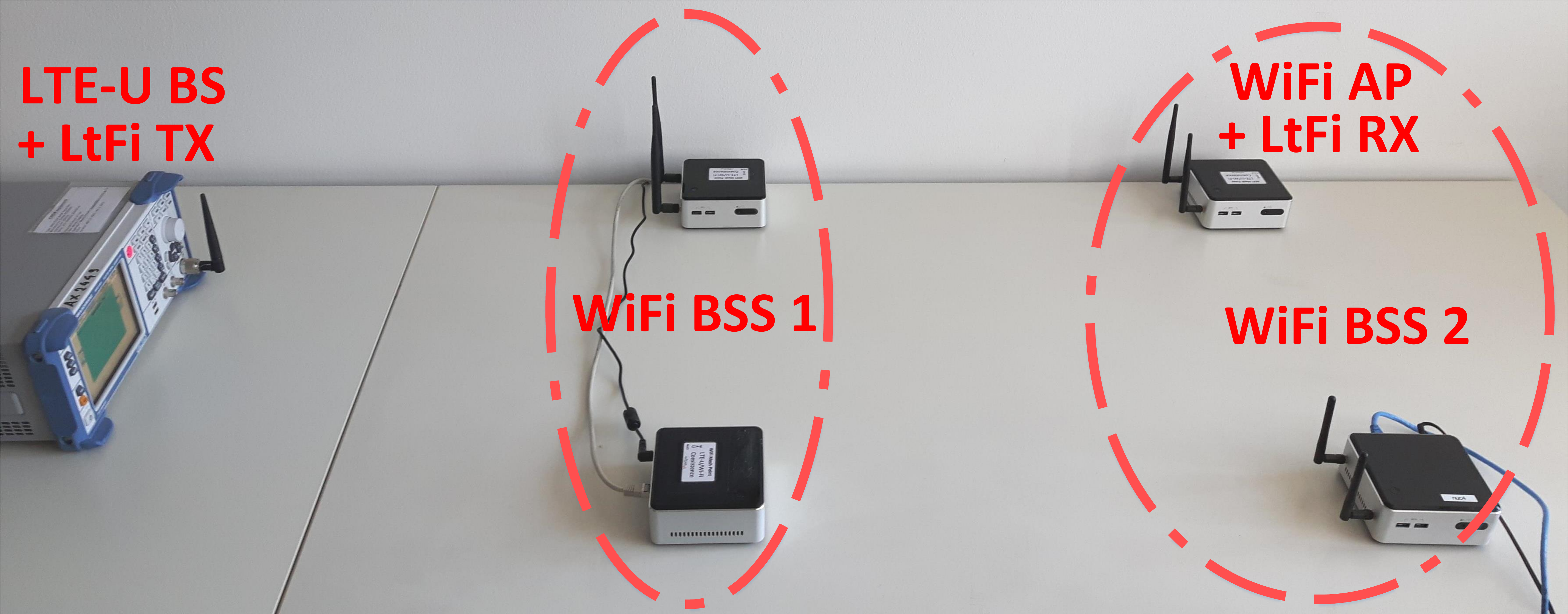}
	\vspace{-20pt}
	\caption{Experiment setup --- LTE-U BS (left) with co-located two WiFi BSSs (right).}
	\label{fig:exp_setup}
\end{wrapfigure}
}%
{
\begin{figure}[ht]
	\centering
	\includegraphics[width=1\linewidth]{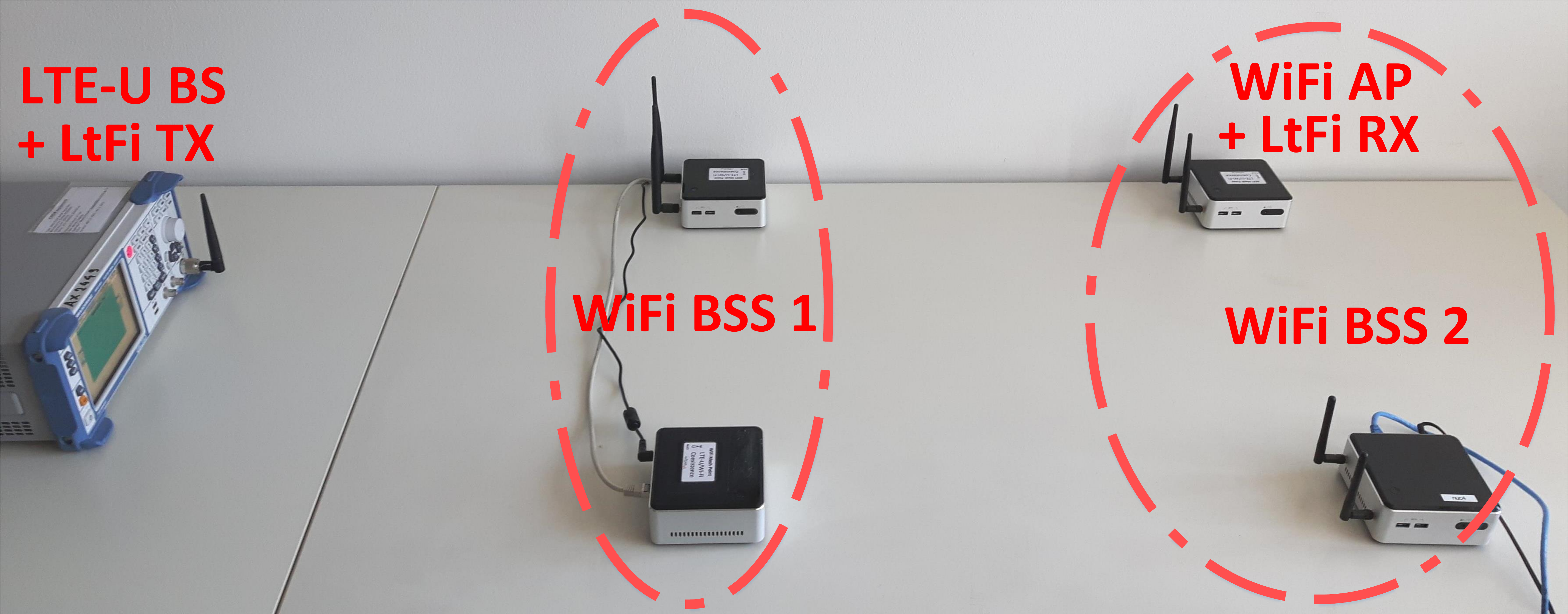}
	\vspace{-20pt}
	\caption{Experiment setup --- LTE-U BS (left) with co-located two WiFi BSSs (right).}
	\label{fig:exp_setup}
	\vspace{-10pt}
\end{figure}
}

The experiment setup is shown in Fig.~\ref{fig:exp_setup} and consists of a single LTE-U BS (R\&S signal generator) and two WiFi BSSs (AP with associated STA). LtFi was running on one of the WiFi APs which was placed 2\,m away from the LTE-U BS. During the experiment the transmission power of the LTE-U BS, i.e. LtFi TX, was varied. The LTE-U CSAT period and \textit{ON-phase} duration was set to 40\,ms and 12\,ms, respectively. With such configuration, LtFi achieves a transmission rate of 75\,bps over the air.

As performance metric for the LtFi air-interface we identified the Frame Error Rate (FER) and Symbol Error Rate (SER). We measured both values for different received signal strength levels of LTE-U signal at the LtFi-enabled WiFi node in four different scenarios, namely:
\begin{enumerate}
\item \textbf{Clear channel:} the wireless channel was free from WiFi traffic and only the LTE-U BS was transmitting during its ON-time. The four WiFi nodes were idle.
\item \textbf{Background traffic:} similar to scenario 1, except that the non-LtFi WiFi AP was generating WiFi background traffic by transmitting DL data to its STAs. Here we distinguished between two cases, namely i) lightly (UDP 10\,Mbit/s) and ii) highly (backlogged TCP traffic) loaded traffic.
\item \textbf{WiFi AP DL:} similar to scenario 1, except that the LtFi-enabled WiFi AP itself was sending DL traffic to its client station. As in scenario 2 we have two cases with light and high load.
\end{enumerate}

The \textit{clear channel} scenario represents the simplest environment for LtFi due to absence of any other signal except the LTE-U. The \textit{background traffic} is more challenging as the LtFi RX node receives a mix containing LTE-U and WiFi signals. The last scenario, \textit{WiFi AP DL} is the worst case as here the LtFi node itself is transmitting WiFi traffic. Due to the half-duplex constraint of WiFi transceiver the LtFi node is unable to receive the LtFi signal while it is transmitting WiFi traffic. 

\subsection{Results}

Fig.~\ref{fig:ltfi_fer} shows the results of the FER for the three different scenarios. We can clearly see that the LtFi air-interface is able to operate close to the receive signal strength required for energy detection based carrier sensing, e.g. in \textit{clear channel} scenario a power level of -60.5\,dBm is sufficient to reliably decode LtFi frames. For the other two scenarios a slightly higher receive power is required, i.e. up to -57\,dBm for \textit{background (high)}. Moreover, we see a very narrow region with intermediate FERs, i.e. 1-2\,dB for \textit{clear channel}. Furthermore, we can see a 1\,dB loss for the high load case. Finally, we see that in \textit{WiFi AP DL (high)} the FER stays above 20\,\% even for high receive power levels. This can be explained by the mentioned half-duplex constraint. The SER is shown in Fig.~\ref{fig:ltfi_ser}. Interestingly, here the SER is smallest in \textit{background (light)}. 

\iftoggle{techreport}{
\begin{figure}
    \centering
    \begin{minipage}{.48\textwidth}
      \centering
      \includegraphics[width=1\linewidth]{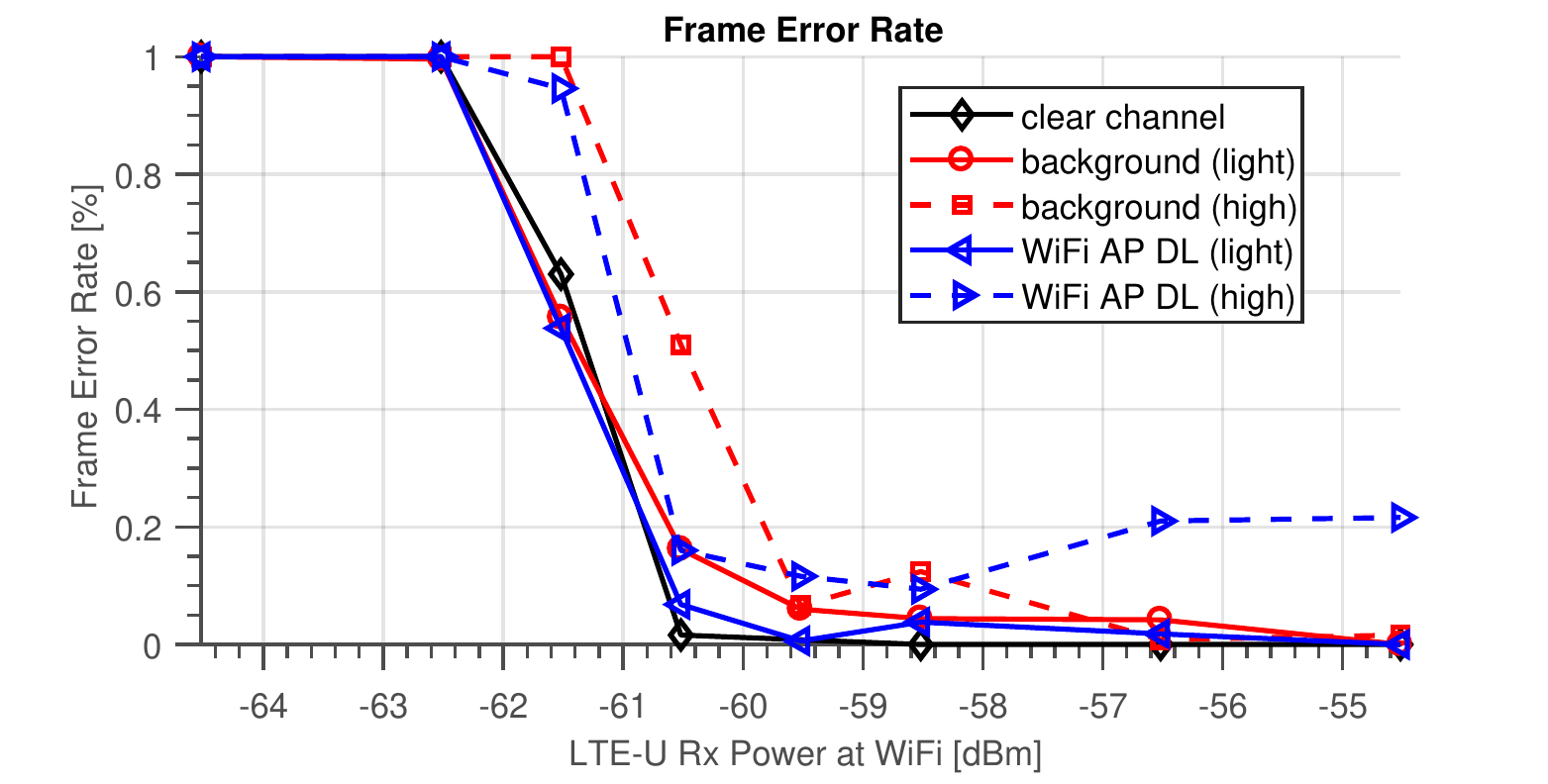}
      \caption{LtFi demodulator performance --- Frame Error Rate (FER) vs. LTE-U RX power with default ED threshold.}
      \label{fig:ltfi_fer}
    \end{minipage}%
    \hfill
    \begin{minipage}{.48\textwidth}
      \centering
      \includegraphics[width=1\linewidth]{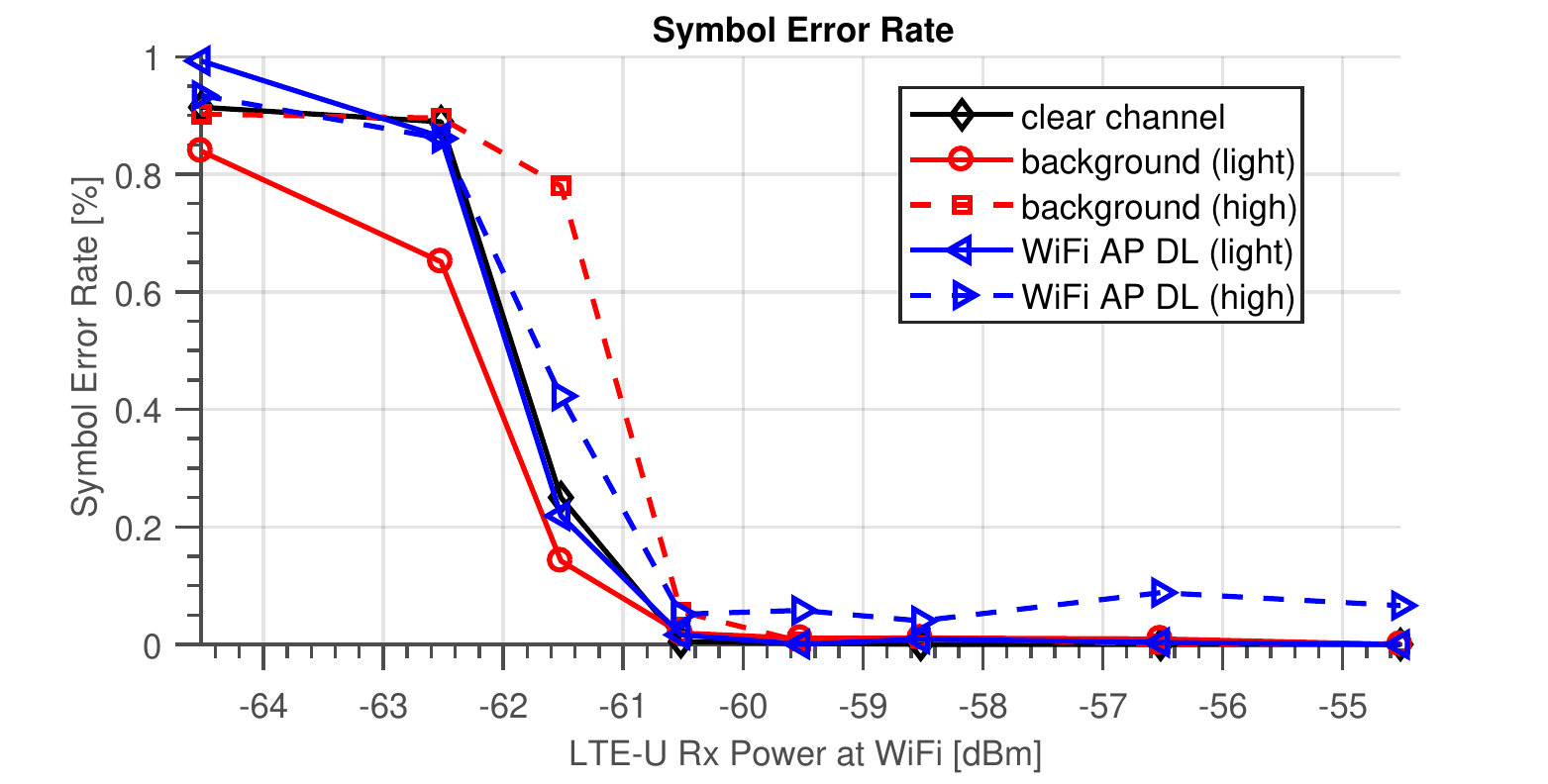}
      \caption{LtFi demodulator performance --- Symbol Error Rate (SER) vs. LTE-U RX power with default ED threshold.}
      \label{fig:ltfi_ser}
    \end{minipage}
\end{figure}
}%
{
\begin{figure}[!h]
	\centering
	\includegraphics[width=1\linewidth]{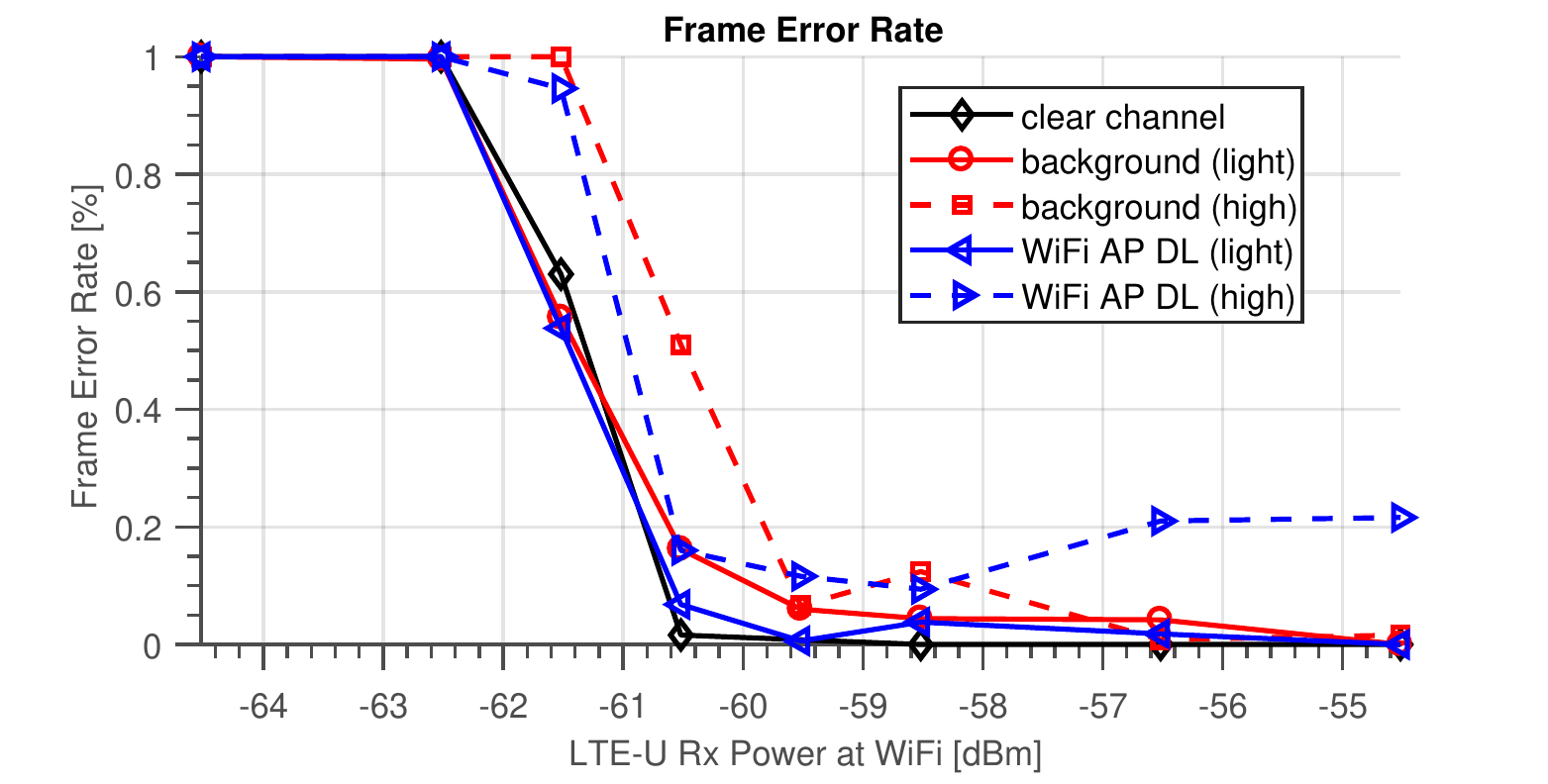}
	\vspace{-20pt}
	\caption{LtFi demodulator performance --- Frame Error Rate (FER) vs. LTE-U RX power with default ED threshold.}
	\label{fig:ltfi_fer}
	\vspace{-10pt}
\end{figure}

\begin{figure}[!h]
	\centering
	\includegraphics[width=1\linewidth]{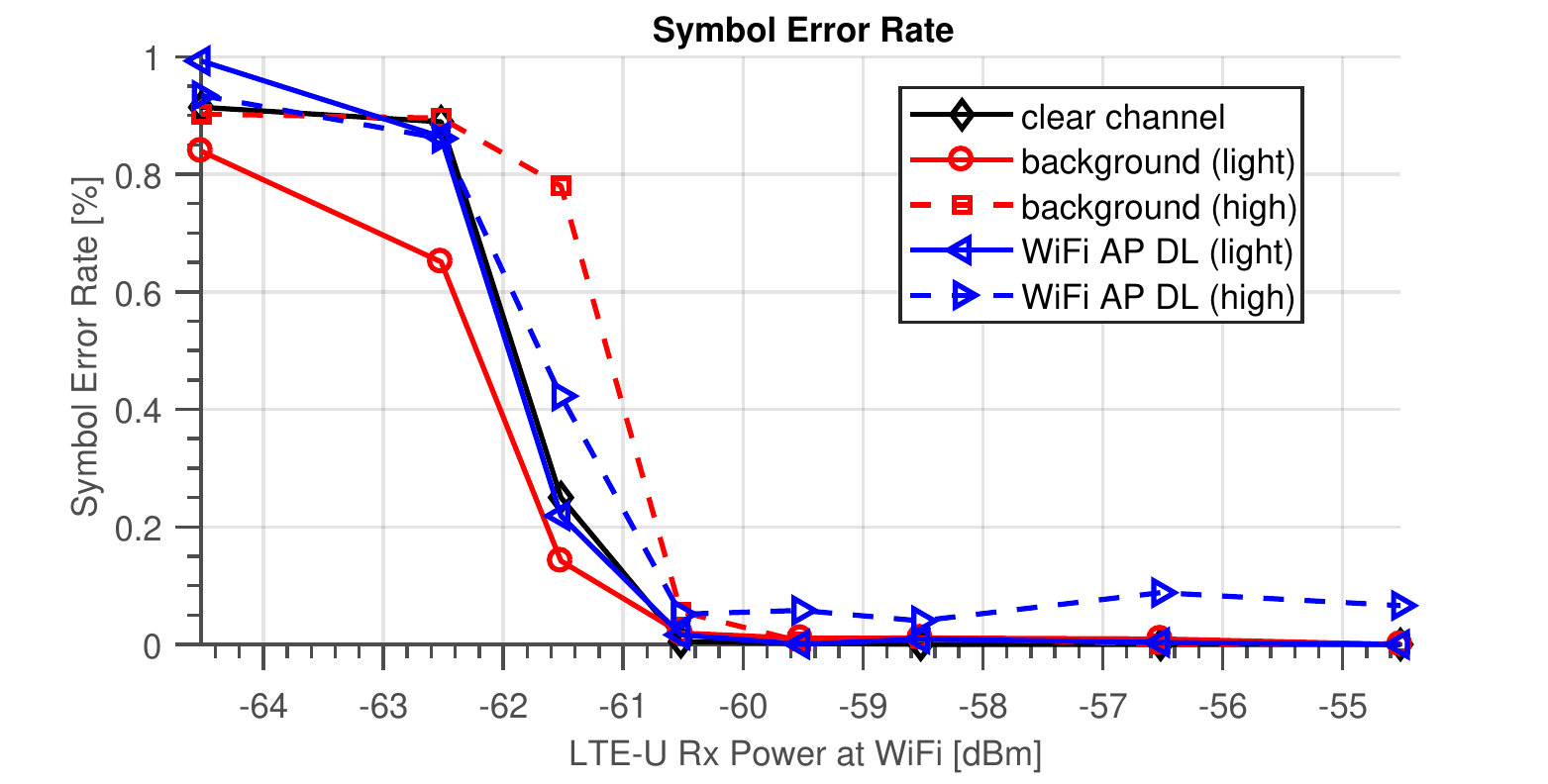}
	\vspace{-20pt}
	\caption{LtFi demodulator performance --- Symbol Error Rate (SER) vs. LTE-U RX power with default ED threshold.}
	\label{fig:ltfi_ser}
\end{figure}
}

\iftoggle{techreport}{
\begin{wrapfigure}[15]{R}{0.5\textwidth}
	\centering
	\includegraphics[width=1\linewidth]{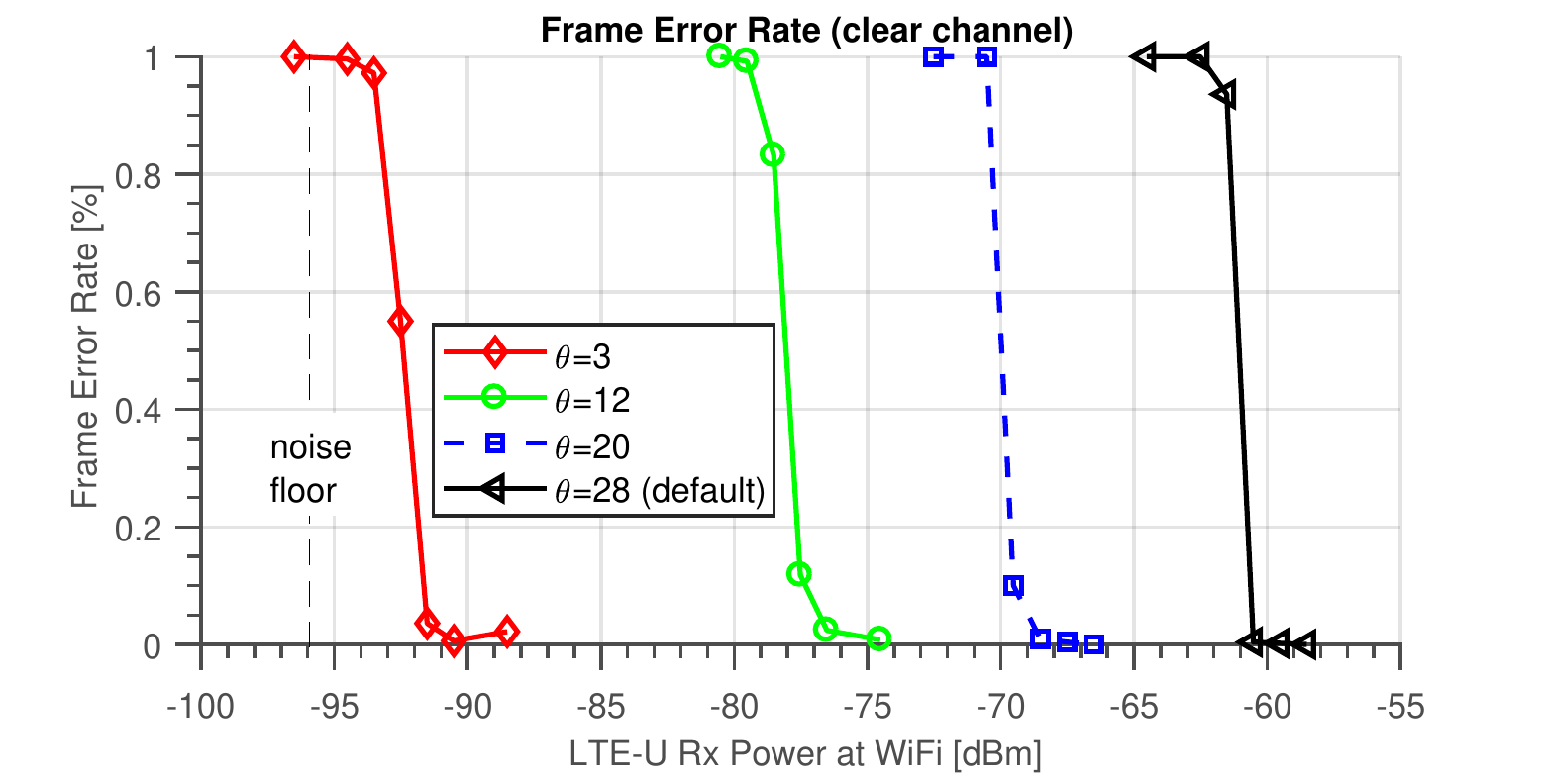}
	\vspace{-20pt}
	\caption{LtFi demodulator performance --- Frame Error Rate (FER) vs. LTE-U RX power using different ED thresholds (Atheros NIC).}
	\label{fig:ltfi_fer_vared}
	\vspace{0pt}
\end{wrapfigure}
}%
{
\begin{figure}[!ht]
	\centering
	\includegraphics[width=1\linewidth]{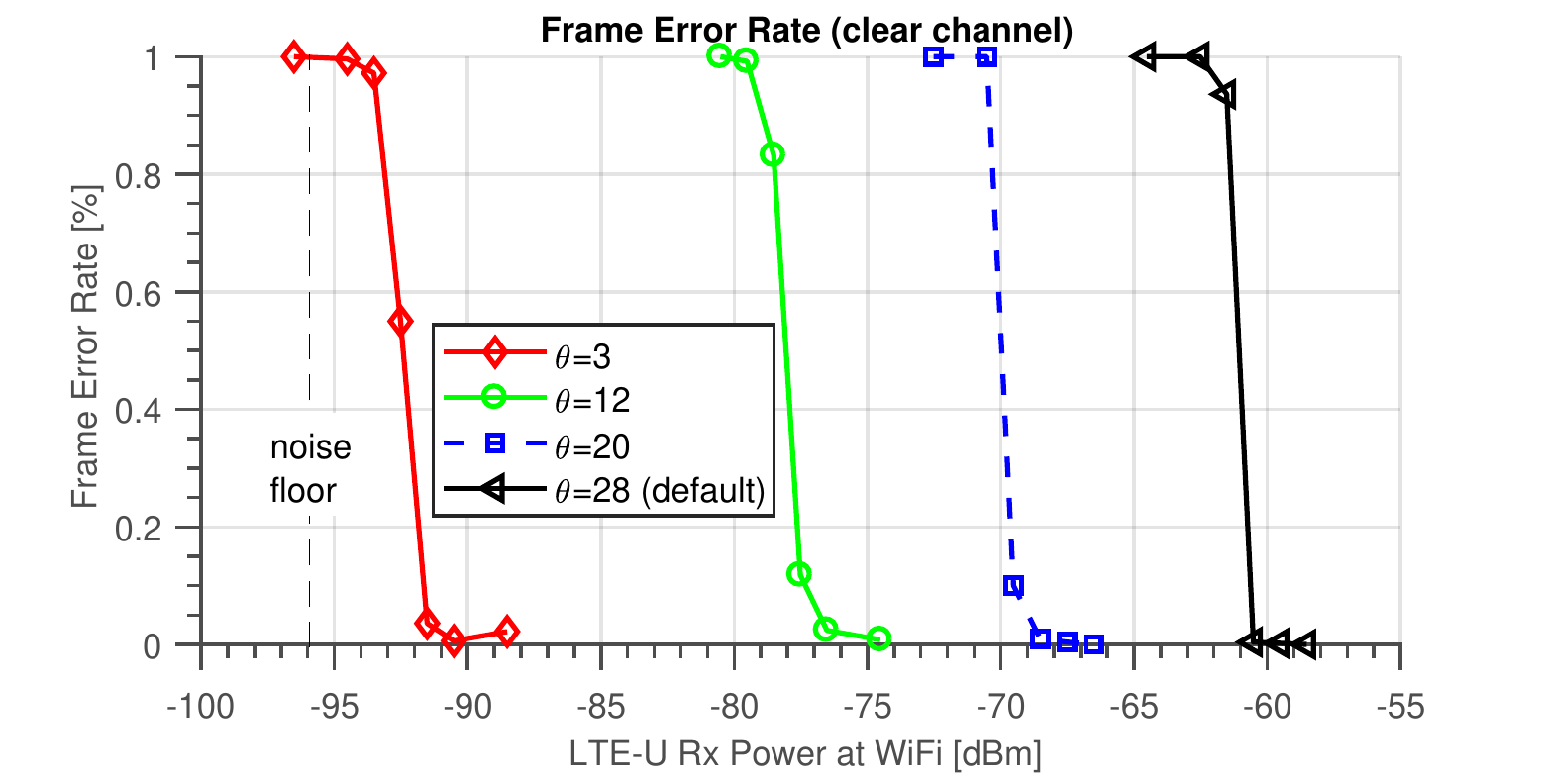}
	\vspace{-20pt}
	\caption{LtFi demodulator performance --- Frame Error Rate (FER) vs. LTE-U RX power using different ED thresholds (Atheros NIC).}
	\label{fig:ltfi_fer_vared}
	\vspace{0pt}
\end{figure}
}

So far we kept the Energy Detection (ED) thresholds of the LtFi RX node constant at its default configuration, i.e. as set by the ATH9k driver. Next, Fig.~\ref{fig:ltfi_fer_vared} shows the FER for the clear channel scenario for different ED values\footnote{Atheros chips allows for changing ED threshold by writing its value to \texttt{AR\_PHY\_CCA} register.}. Note, the black curve ($\theta=28$) corresponds to the default configuration used in the Atheros WiFi NIC. The highest sensitivity we achieved with $\theta=3$ where LtFi is able to decode the signal at very low receive power levels, i.e. -92\,dBm.

\medskip
\noindent \textit{\textbf{Takeaways: }}The information sent of the LtFi air-interface can be reliably decoded at the LtFi receiver at even very low LTE-U receive power levels. With the default ED configuration used by the WiFi NIC it is around -61\,dBm. By adapting the ED threshold the sensitivity can be dramatically increased to around -92\,dBm.


%
%
\section{System-level Simulations}

We evaluated LtFi system-wide. The objective was to show that from the LtFi's air-interface a WiFi node is able to estimate the set of LTE-U BSs in its proximity. Therefore we consider a typical LTE-U BS deployment, i.e. hexagonal placement of BSs with frequency-reuse 1 and omni-directional antennas.

\subsection{Methodology}

We conducted system-level simulations using Matlab according to the methodology recommended by the IEEE 802.16m group~\cite{zhuang09emd}. The setup mimics an indoor small office scenario. The LTE-U inter-BS distance was set to 50\,m. All LTE-U BSs are using the same unlicensed channel (5.2\,GHz) and are transmitting the LtFi signal as described in Sec.~\ref{sec:multicell}. For the simulations we used the following simplified model for the LtFi receiver according to which the LtFi receiver is able to perfectly decode the information received over the LtFi air-interface as long as the wanted CTC signal was above and the interfering CTC signal was below the sensitivity level of -77\,dBm, i.e. 19\,dB above the noise floor, respectively. The value was selected as it achieves the best performance\footnote{In case of Atheros WiFi NIC the ED is configured with $\theta=23$.}. Finally, the LtFi receiver was placed on a regular grid in the bounding box with side length of 140\,m. At each location the LtFi proximity detection algorithm (Listing~\ref{proximity_algo}) was executed in order to estimate the number of LTE-U BSs in its vicinity. The remaining parameters are summarized in Table~\ref{table:simparams}.

\begin{table}[t]
\footnotesize
\centering
\caption{Simulation parameters.}
\vspace{-5pt}
\begin{tabular}{p{.4\linewidth}p{.4\linewidth}}
	\colheadbegin \textbf{Parameter} & \textbf{Value} \colheadend
	LTE-U TX power/FFR/MAC& 20\,dBm/1/CSAT\\
	WiFi noise figure& 6\,dB\\
	LtFi RX sensitivity rel. to noise & 17\,dB (-77\,dBm)\\
	Pathloss model& Motley-Keenan ($\alpha = 0.44$)\\
	Correlated shadowing $\sigma$& 0 \& 6\,dB\\
	LTE-U placement& 100 BSs placed in hexagonal\\
	LTE-U inter-BS distance& 50\,m\\
\end{tabular}
\label{table:simparams}
\vspace{-10pt}
\end{table}

\subsection{Results}

Fig.~\ref{fig:sim_res_no_shadow} shows for each LtFi receiver location (point in space) the estimated number of LTE-U BSs. In absence of Shadowing, i.e. $\sigma = 0$, we can observe a strong correlation between the LtFi's receiver positions and the number of estimated LTE-U BSs. For location very close to LTE-U BSs the number of reported BSs is up to seven whereas for locations between three BSs the reported number is three.

Finally, Fig.~\ref{fig:sim_res_shadow} shows the results for an environment with Shadowing, i.e. $\sigma = 6$.

\medskip
\noindent \textit{\textbf{Takeaways: }}In a LTE-U network where BSs are operating on the same unlicensed channel LtFi is able to accurately estimate the BSs in its proximity.

\iftoggle{techreport}{

\begin{figure}[!tbp]
  \centering
  \subfloat[Without shadowing.]{\includegraphics[width=0.45\linewidth]{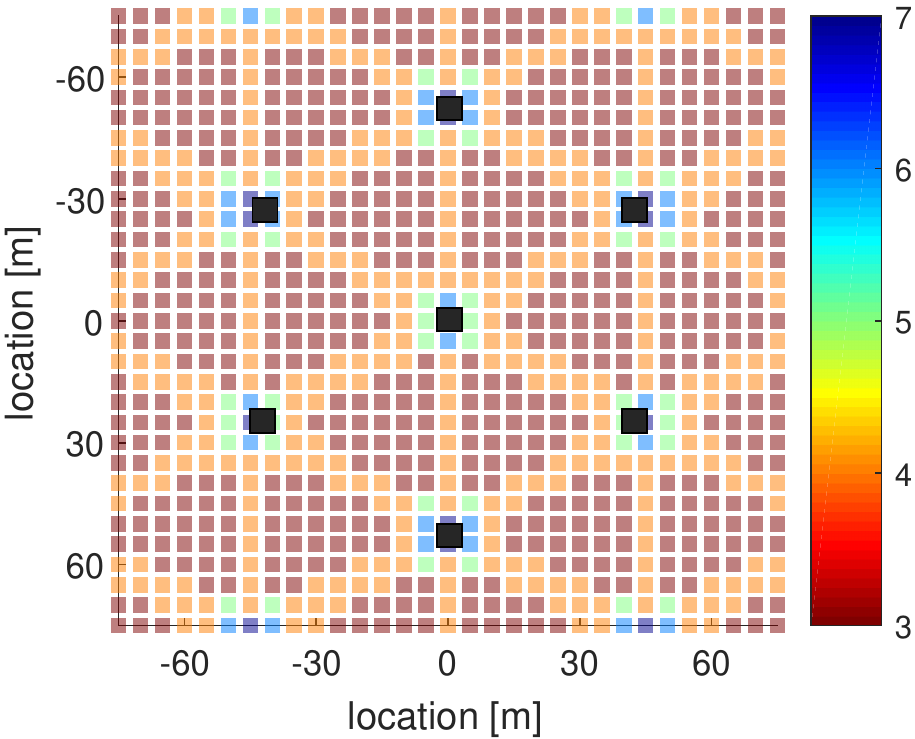} \label{fig:sim_res_no_shadow}}
  \hfill
  \subfloat[With Shadowing.]{\includegraphics[width=0.45\linewidth]{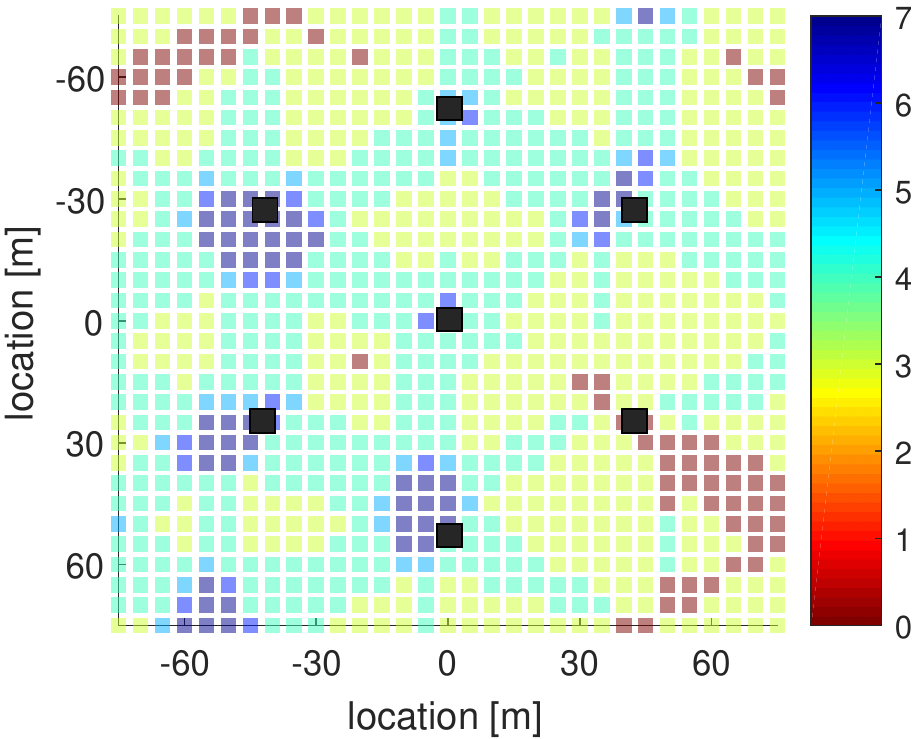}\label{fig:sim_res_shadow}}
\caption{Number of detected LTE-U BSs at each spatial location. Black rectangles mark the location of LTE-U BSs. An ED threshold of $\theta=23$ (-77\,dBm) is used.}
\end{figure}
}%
{
\begin{figure}[!t]
    \centering
    \begin{minipage}[b]{1\linewidth}
    \centering
    \subfloat[][\label{fig:sim_res_no_shadow}Without shadowing]{
    \includegraphics[width=0.8\linewidth]{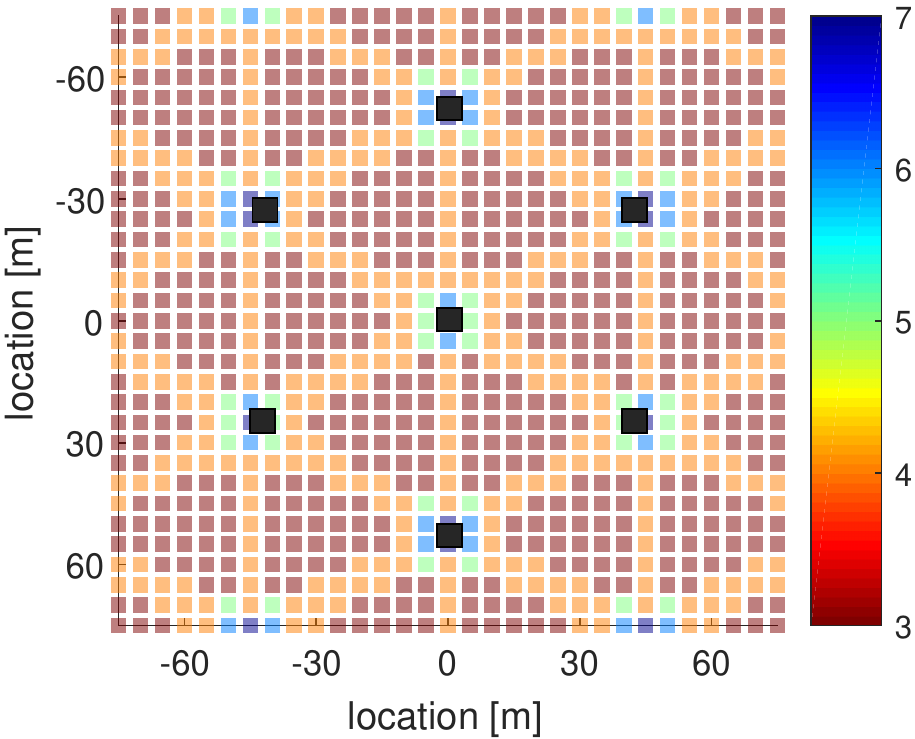}
    }\\
    \subfloat[][\label{fig:sim_res_shadow}With Shadowing]{
    \includegraphics[width=0.8\linewidth]{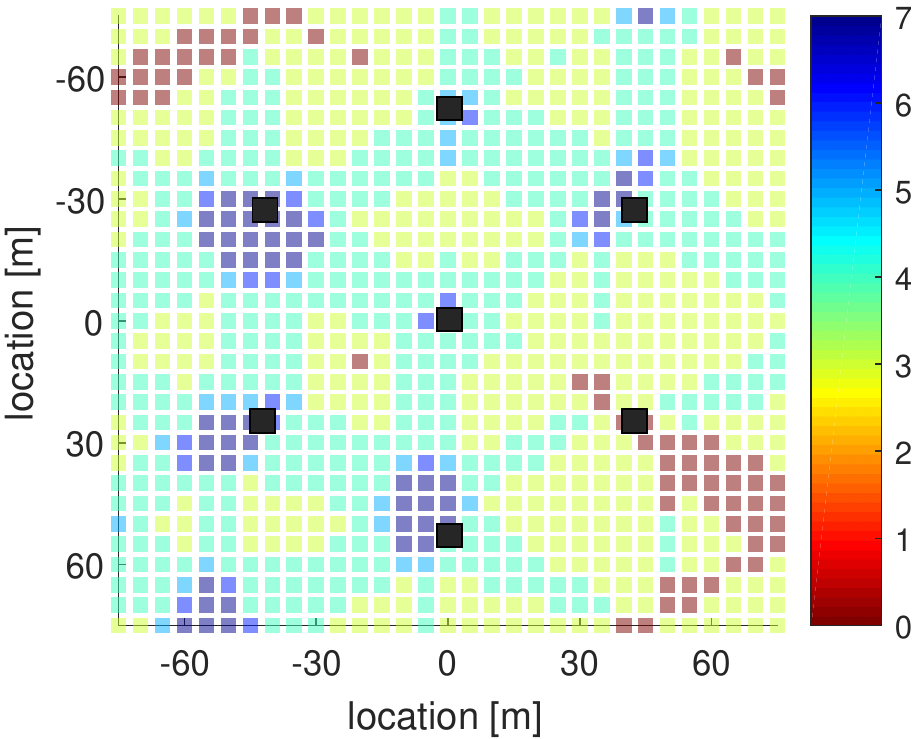}
    }
    \caption{\label{fig:sim_res}
    Number of detected LTE-U BSs at each spatial location. Black rectangles mark the location of LTE-U BSs. An ED threshold of $\theta=23$ (-77\,dBm) is used.
    }
    \end{minipage}
    \vspace{-15pt}
\end{figure}
}


\section{Related Work} \label{chapter:related_work}

The present work is based on our previous work where we proposed an architecture for setting up out-of-band control channel between homogeneous network nodes, namely residential WiFi APs~\cite{Zehl16resfi}. With LtFi we extended this idea towards cross-technology control between heterogeneous network nodes, here LTE-U and WiFi. To the best of our knowledge, there is no previous work on cross-technology communication between LTE-U and WiFi.

So far, the research focus was to enable cross-technology communication between WiFi and sensor networks (mostly ZigBee), that coexist in the same 2.4GHz band.
Esense~\cite{Esense} and HoWiES~\cite{HoWiES} enable over-the-air WiFi to ZigBee communication by injecting dummy packets with durations that are unlikely to be used in normal WiFi traffic. They can achieve relatively high throughput, but are burden to already saturated spectrum. 
GapSence~\cite{GapSense} prepends legacy packets with a customized preamble containing sequences of energy pulses. The length of silent gaps between them encodes the CTC data to be transmitted. Such approach requires a dedicated hardware and is not compatible with COTS devices.
FreeBee~\cite{FreeBee} modulates CTC data by shifting the timing of periodic beacon frames, but suffers from low rate being limited by the beacon rate.
C-Morse~\cite{cmorse}, DCTC~\cite{TransparentCTC}, EMF~\cite{emf} and WiZig~\cite{wizig} achieve high CTC rates by utilizing all types of frames. 
In general, they slightly perturb the transmission timing of WiFi frames to construct recognizable radio patterns within negligible delay. Furthermore, they are compliant with existing standards and strive to be transparent to upper protocol layers. In contrast, in case of LTE-U we cannot modify the transmission timing, as it is tightly scheduled.

%
%



\section{Conclusions}

In this paper we introduced LtFi, a system which enables to set-up a CTC between LTE-U and WiFi. LtFi is fully compliant and transparent with LTE-U technology and works with WiFi COTS hardware. LtFi is of low complexity and fully compliant (transparent) with LTE-U technology. It requires just a simple interface to the LTE-U eNb scheduler to program the data to be transmitted over the LtFi air interface. On WiFi side LtFi needs an interface for sampling the MAC state (registers) which is already provided by COTS NICs like Atheros 802.11n/ac. The LtFi X2 interface creates a bi-directional CTC over the Internet which enables to perform cross-technology interference and radio resource management.

For future work, we plan to replace our LTE-U hardware by a more flexible system composed of LTE-U implemented in srsLTE~\cite{srsLte} or Eurecom’s OpenAirInterface (OAI)~\cite{openAir} on Software Defined Radios like USRP~\cite{ettus} platform. This will allow us to test concrete cross-technology RRM and interference management schemes. 

\section{Acknowledgment}

This work has been supported by the European Union’s Horizon 2020 research and innovation programme under grant agreement No. 645274 (WiSHFUL project).

\bibliographystyle{IEEEtran}
\bibliography{biblio,IEEEabrv}
\end{document}